\documentclass[twoside]{article}

\usepackage[accepted]{aistats2023}
\usepackage[round]{natbib}
\renewcommand{\bibname}{References}

\RequirePackage[OT1]{fontenc}

\RequirePackage{amsthm,amsmath,amsfonts,amssymb,mathtools,thmtools}

\RequirePackage[colorlinks,citecolor=blue,urlcolor=blue]{hyperref}
\RequirePackage{enumitem,graphicx,MnSymbol,mathrsfs}

\usepackage{bm}
\usepackage{graphicx,psfrag,epsf}  
\usepackage{subcaption}
\usepackage{tikz-cd}

\usepackage[title]{appendix}

\theoremstyle{plain}                          
\newtheorem{thm}{Theorem}[section]     

\newcommand{\Hc}{\mathcal{H}}

\definecolor{nRed}{RGB}{220,0,12}
\definecolor{nOrange}{RGB}{254,166,0}
\definecolor{nBlue}{RGB}{40,96,180}

\newcommand{\commentout}[1]{}

\begin{document}
%

%

\twocolumn[

\aistatstitle{Safe Sequential Testing and Effect Estimation in Stratified Count Data}

\aistatsauthor{Rosanne J. Turner \And Peter D. Gr\"unwald }

\aistatsaddress{ CWI, UMC Utrecht \And  CWI, Leiden University } ]

\begin{abstract}
  Sequential decision making significantly speeds up research and is more cost-effective compared to fixed-$n$ methods. We present a method for sequential decision making for stratified count data that retains Type-I error guarantee or false discovery rate under optional stopping, using \emph{e-variables}. We invert the method to construct stratified anytime-valid confidence sequences, where cross-talk between subpopulations in the data can be allowed during data collection to improve power. Finally, we combine information collected in separate subpopulations through pseudo-Bayesian averaging and switching to create effective estimates for the minimal, mean and maximal treatment effects in the subpopulations.
\end{abstract}

\section{INTRODUCTION}
Fixed-n hypothesis tests and confidence intervals limit research opportunities and quick decision making, as they rely on static research designs where data are only evaluated at one time point. We aim to develop hypothesis tests for conditional independence and anytime-valid confidence sequences for stratified treatment effects in subpopulations that retain a guarantee on the probability of falsely rejecting the null hypothesis and coverage of the true effect under continuous monitoring of data. 
To this end we use {\em e-values}, tools for constructing tests that keep the \emph{type-I error rate} (or \emph{false positive rate}) controlled under sequential testing with optional stopping. Over the last four years, e-values have become the standard tools (essentially, the appropriate alternative for $p$-values) for dealing with such settings. Below we summarize the essentials; for much more background on the budding field of e-processes (also known as `testing by betting' and `safe testing') see the recent overview \citep{ramdas2022game} and specifically for details on e-values refer to \cite{grunwald2019safe,VovkW21}. In this paper, we develop e-processes for stratified $2 \times 2$ tables, enabling, in Section~\ref{sec:global}, anytime-valid (i.e. valid under optional stopping) conditional independence (CI) tests for Bernoulli streams for two groups $a$ and $b$ (e.g. $a$ is control, $b$ is treatment), where the test is conditional on a third variable, the stratum. Based on these CI tests, we then, in Section~\ref{sec:confidence}, develop anytime-valid confidence sequences (henceforth just called `confidence sequences') for a notion of effect size representing divergence from CI. The importance of our tests is ubiquitous in e.g. medical statistics --- we can think of the CI test in Section~\ref{sec:global} as an an anytime-valid sequential version of the Cochran-Mantel-Haenzel test, a work-horse in the field of  epidemiology. 
Our e-processes are generalizations of those designed for $2 \times 2$ tables (same setting as ours, but with just a single stratum) by \cite{turner2021safe,turner2022confseq}. To achieve the generalization, we employ tools from the theoretical machine learning literature, most notably the literature on {\em prediction with expert advice} \citep{CesaBianchiL06}, which extends Bayesian learning techniques with ideas such as `sleeping', `switching' and the like. Moreover, inspired by these ideas, we develop the novel notion of {\em cross-talk\/} between strata, which allows us to make confidence intervals {\em narrower\/} if outcomes in various strata are interrelated, while nevertheless remaining {\em valid\/} even if they are not. 
While for many statistical models, anytime-valid tests need more data to reach a desired conclusion than fixed $n$ methods and anytime-valid confidence intervals are somewhat wider than standard ones \citep{ramdas2022game,grunwald2019safe}, we find in this paper that we can partially counteract this difference by employing the cross-talk strategy (which is not available for fixed-$n$ methods), as is illustrated by comparing our confidence sequences to fixed-$n$ confidence intervals for Mantel-Haenszel risk differences in  Section~\ref{sec:confidence}. 
\paragraph{E-Processes} Consider a random process  $Y_1, Y_2, \ldots$ and let $\Hc_0$, the {\em null hypothesis}, be a set of distributions for this process. An e-variable for $Y_{j}, Y_{j+1}, \ldots, Y_m$ conditional on $Y^{(j-1)} = (Y_1, \ldots, Y_{j-1})$ for testing $\Hc_0$ is any nonnegative random variable $S$ that can be written as  function of $Y^{(m)}= (Y_1, \ldots, Y_{m})$ such that 
\begin{equation}\label{eq:evar}
    \forall P \in \Hc_0: \mathbb{E}_{P}[S \mid S^{(j-1)}] \leq 1;
\end{equation}
for $j=1$ we set  $\mathbb{E}_P[S \mid S^{(0)}] := \mathbb{E}_P[S]$ and call $S$ an {\em unconditional\/} e-variable; an e-value is the value an e-variable takes on a realized sample. It is easily shown that for  any sequence $S_1, S_2, \ldots$ where $S_j$ is an e-variable for $Y_{(j)}$ conditional on $Y^{(j-1)}$, the product $E^{(m)} := \prod_{j=1}^m S_{j}$ is an unconditional e-variable for $Y^{(m)}$.  $E^{(1)}, E^{(2)}, \ldots$ is called a {\em test martingale\/} or {\em e-process} (see \cite{ramdas2022game} on how e-processes strictly generalize test martingales). Via Ville's inequality, it is shown that e-processes have the remarkable property that, for any $0 < \alpha< 1$, {\em the probability that there exists an $m$ such that $E^{(m)} \geq 1/\alpha$ is bounded by $\alpha$}. As a consequence, if we look at the data at some time $m$ and reject if $E^{(m)} \geq 1/\alpha$, the probability under the null of falsely rejecting the null is at most $\alpha$ no matter how we chose $m$; it may be determined by external circumstances (do we have money to experiment further?) or by aggressive stopping rules such as `keep sampling until you can reject the null', or even by peeking into the future.
Tests with this property are called {\em safe under optional stopping\/} and \cite{ramdas2020admissible} show that, in essence, all reasonable such tests should be based on e-processes. Just like p-values can be converted into confidence intervals, e-process can be converted into anytime-valid confidence tests, also known as {\em confidence sequences} --- we will explore these in Section~\ref{sec:confidence}. 
\paragraph{Setting}
We consider the \emph{stratified contingency table} setting/model. 
Under the {\em global null\/} hypothesis (we consider more complicated nulls later), outcomes $Y \in \{ 0, 1\}$ are independent of
groups $X \in \{a, b\}$ (e.g. representing interventions) given their stratum $k \in [K]:= \{1, ..., K\}$. We formalize this by  measuring time in terms of {\em blocks\/}: 
we assume that at each time $j=1,2,\ldots$, we are given a stratum indicator  $k_j \in [K]$ and we observe a block of $n = n_a+n_b$ outcomes, with $n_a$ outcomes in group $a$ and $n_b$ in group $b$, all in the same stratum $k_j$. 
We write  $Y^{(m)} = (Y_{1}, \ldots, Y_{m})$ with $Y_{j}$ the data vector corresponding to the $j$-th block arriving. Hence $Y_{j} = (Y_{j,a,1}, \ldots, Y_{j,a,n_a},Y_{j,b,1}, \ldots, Y_{j,b, n_b})$  is a vector in $\{0,1\}^{n}$ denoting $n=n_a+n_b$ outcomes in $k_j$. 
Under both null and alternative, all blocks are assumed independent, with each outcome in group $x$ in stratum $k$ independently $\sim \text{Bernoulli}(\theta_{x,k})$. Formally, the null hypothesis then expresses that 
\begin{equation}
\label{eq:generalH0}
    \Hc_0: \theta_{a,k} = \theta_{b,k} \text{ for all } k.
\end{equation}

We will assume $n_a = n_b = 1$ for all strata in simulation examples in this paper, but these can be chosen freely in practice and can even be adapted in between data blocks --- as long as they are set at or before the beginning of a data block, they are allowed to depend on the past. 
Of course,  in practice, we often deal with $2K$ i.i.d. streams of data, one for each group-stratum combination, with data not necessarily coming in at the same rate for different strata/groups. While superficially different, we can still recast this setting in terms of blocks: for example, participant may sequentially enter  a study and are each independently  randomized with probability $1/2$ to receive `treatment' (group $b$) or `control/placebo' ($a$). We then wait until the first time $t_1$ that we have seen $n_a$ outcomes in group $a$ and $n_b$ outcomes in group $b$ in the same stratum; we call this stratum $k_1$, denote these $n$ outcomes $Y_1$, and  proceed observing outcomes in the various streams  until the first time $t_2$ that there is another stratum $k_2$ (potentially $k_2=k_1$) so that we have seen $n_a$ outcomes in group $a$, $n_b$ in group $b$ in stratum $k_2$; we denote these $n$ outcomes $Y_2$, and so on. If we want to stop at any time $t$, we take as data all blocks that have been completed so far, and ignore all started-yet-unfinished blocks. 

\paragraph{Related Work} 
The first paper to use  e-processes for conditional independence testing  is \citep{lindon2020anytime}, but their tests are very different from ours and involve a {\em simple\/} null hypothesis, allowing them to use Bayes factors for their e-processes. Further,  \cite{turner2021safe,turner2022confseq} develop independence tests and confidence sequence for $2 \times 2$ tables; our paper is a direct extension of theirs, extending their techniques to the stratum-conditional case. Very recently four other related papers have appeared:  \citep{pandeva2022,GrunwaldHL22,shaer2022model,Duan2022}: these papers all differ from ours in that they assume data are jointly  i.i.d. (i.e. one observes a single i.i.d. stream $(X_1, Y_1, K_1), (X_2, Y_2, K_2), \ldots $). The latter three also make the so-called {\em Model-X\/} assumption (the distribution of $X_i \mid K_i$ is assumed known). Our paper is complementary: we do not need the i.i.d. or Model-X assumption and as explained above, our setting does not just capture data in blocks (such as paired data) but also data in the form of $2K$ i.i.d. streams, one for each group in each stratum, with no stochastic assumptions about what group or what stratum arrives at what time. The price we have to pay is that we can only deal with a small number of strata and with finite sets of outcomes and number of groups (in this paper we focus on $2$ but extension to the finite case is straightforward); aforemetioned references can deal with arbitrary covariate and outcome random variables $K_i$ and $Y_i$. 
Nevertheless, small-strata-count-studies are highly common in the medical statistics world, and we show here how to construct efficient sequential tests for them. 

The code used for experiments in this paper will initially be placed on the repository linked to this publication \citep{turner2023safesequential}, and will later be integrated in the safestats R package \citep{ly2022safestats}.

%
%

\section{E-VARIABLES FOR TESTING THE GLOBAL NULL}\label{sec:global}
We first consider the case where there is only one stratum, $k_j=k^*$ for each each $j$. The problem is then reduced to testing whether two Bernoulli data streams come from the same source. \cite{turner2021safe} showed that in this case, for arbitrary estimators $\breve\theta_a|Y^{(j-1)},\breve\theta_b|Y^{(j-1)}$, the following is an e-variable for $Y_j$ conditional on $Y^{(j-1)}$, i.e.  (\ref{eq:evar}) holds with $S := S_j$ given by 
\begin{align}\label{eq:niceEvar}
S_j = 
\prod_{i=1}^{n_a} 
\frac{p_{\breve\theta_a|Y^{(j-1)}}(Y_{j,a,i})}{
p_{\breve\theta_0 |Y^{(j-1)}}(Y_{j,a,i}
)} 
\prod_{i=1}^{n_b} 
\frac{p_{\breve\theta_b|Y^{(j-1)}}(Y_{j,b,i})}{
p_{\breve\theta_0 |Y^{(j-1)}}(Y_{j,b,i})
}, 
\end{align}
where $p_{\theta}(Y) = \theta^Y (1-\theta)^{1-Y}$ denotes the Bernoulli$(\theta)$ probability of $Y \in \{0,1\}$), as long as we pick $\breve\theta_0 
\in \Theta_0 = [0,1]$ as follows: 
\begin{align}\label{eq:kl}
    \breve\theta_0 = \breve\theta_0 |Y^{(j-1)}  & := 
    \arg \min_{\theta \in [0,1]} D(P_{\breve\theta_a,\breve\theta_b} \| P_{\theta,\theta}) 
   \nonumber  \\ & 
    \overset{(a)}{=}
    \frac{n_a}{n}
\breve\theta_a| Y^{(j-1)} + \frac{n_b}{n}\breve\theta_b | Y^{(j-1)}.
\end{align}
Here and in the sequel, $P_{\theta_a,\theta_b}$ represents the  distribution on $n_a+ n_b$ independent binary outcomes with the first $n_a$ outcomes $\sim$ Bernoulli$(\theta_a)$ and the subsequent $n_b$ outcomes $\sim$ Bernoulli$(\theta_b)$, i.e. the distribution of outcomes in a single block according to $(\theta_a,\theta_b)$, and 
$D(P_{\theta_a,\theta_b} \| P_{\theta'_a,\theta'_b})$
abbreviates the KL divergence between two such distributions. 
Equality (a) follows by simple calculus. 

Importantly, in (\ref{eq:niceEvar}), 
\emph{$(\breve\theta_a,\breve\theta_b) \in \Theta_1 = [0,1]^2$ can be chosen as a function of past data anyway we like, not affecting the Type-I error guarantee}. Nevertheless, if we were given the true probabilities $\theta^*_a$ and $\theta^*_b$ of the two groups in block $j$, then we could set  $\breve\theta_a = \theta^*_a$ and $\breve\theta_b = \theta^*_b$ and this choice is  special: the e-variable \eqref{eq:niceEvar} then has, among all e-variables, the largest expected logarithm under the true alternative $P_{\theta^*_a,\theta^*_b}$. We then say it is \emph{growth-rate optimal} (GRO) for collecting evidence against the null hypothesis \citep{grunwald2019safe}. Formally, we define 
\begin{equation}\label{eq:grodef}
\textsc{gro}(\theta^*_a,\theta^*_b) := \sup_{S} {\bf E}_{Y_{j} \sim P_{\theta^*_a,\theta^*_b}}[\log S]
\end{equation}
where the supremum is over all random variables $S$ that are e-variables for $Y_{j}$ under ${\cal H}_0$. 
It directly follows from \cite[Theorem 1]{grunwald2019safe}  that,  if we plug in 
 $\breve\theta_a = \theta^*_a$ and $\breve\theta_b = \theta^*_b$ into (\ref{eq:niceEvar}), then the resulting 
$S_j$ is GRO and its growth rate is equal to the KL divergence, i.e.
\begin{equation}\label{eq:klgro}
 {\bf E}_{Y_{j} \sim P_{\theta^*_a,\theta^*_b}}[\log S_j] = \textsc{gro}(\theta^*_a,\theta^*_b)
= D (P_{\theta^*_a,\theta^*_b} \| P_{\tilde\theta,\tilde\theta} ),
\end{equation} where $\tilde\theta = (n_a/n) \theta^*_a + (n_b/n) \theta^*_b$. 
Growth-rate optimality is the analogue of statistical {\em power\/} in the sequential setting: if we plug in these `true' $\breve\theta_a = \theta^*_a,\breve\theta_b=\theta^*_b$, we expect the product $E^{(m)}$ to increase as fast as possible in $m$, enabling us to reach $1/\alpha$ and reject the null hypothesis as fast as possible, compared with all other possible e-processes.
\commentout{Utilizing one of the special properties of e-Variables, multiplying the subsequent e-Variables in \eqref{eq:niceEvar} yields a test martingale:
\begin{equation}
    \label{eq:testmartingale}
    E^{(m)} = \prod_{j=1}^{m} S_{j}(Y_j),
\end{equation}
a special instance of an \emph{e-process} \citep{ramdas2022testing}. $\Hc_0$ is rejected after block $m$ if $E^{(m)} \geq \frac{1}{\alpha}$, which offers a false positive rate guarantee at level $\alpha$ \citep{grunwald2019safe}. }
In practice though, $\theta^*_a$ and $\theta^*_b$ are unknown, but to get near-grow-optimal e-variables, we can {\em estimate\/}
$\breve\theta_a$ and $\breve\theta_b$ based on all data seen before data block $j$ --- then $\breve\theta_a$ and $\breve\theta_b$ converge to $\theta^*_a,\theta^*_b$ and our e-variables $S_j$ get better and better in the GRO sense. We follow \cite{turner2021safe} who successfully chose to place a beta prior on the parameter space and took the Bayesian posterior mean as an estimate. 

\begin{figure}[ht]
\vspace{.3in}
     \centering
        \includegraphics[width=5cm]{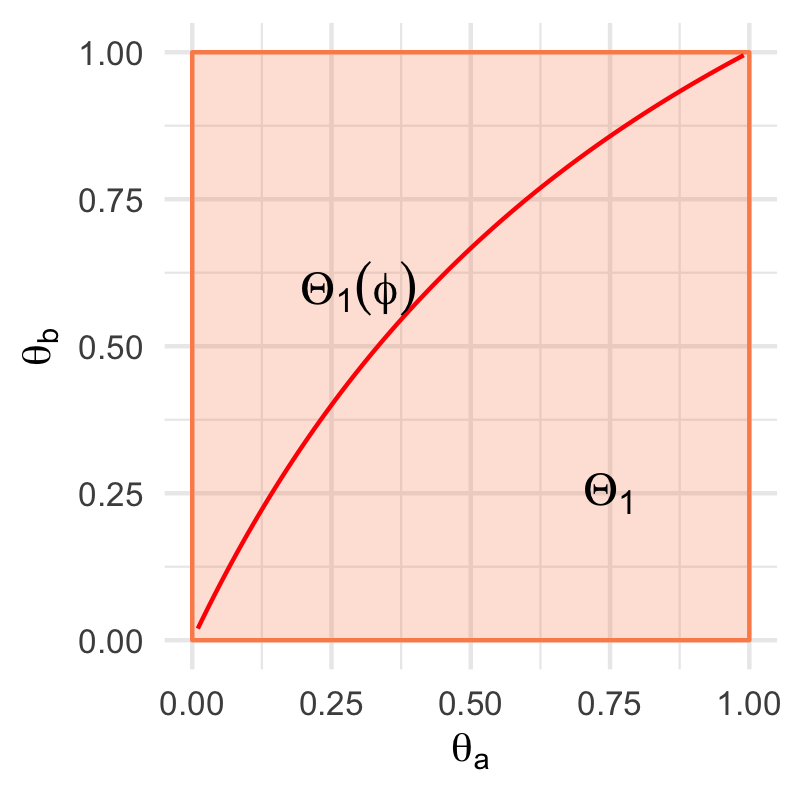}
        \vspace{.3in}
        \caption{Parameter space $\breve\theta_{a}|Y^{(j-1)}$ and $\breve\theta_b|Y^{(j-1)}$ are estimated in, in $2 \times 2$ table without strata; either through placing a beta prior on the entire unit square (in light orange) and calculating the posterior mean with all data up to and including time $j-1$ or through restricting the posterior estimation to a particular odds ratio value $\phi$ and placing a beta prior on all pairs $(\theta_a, \theta_b)$ corresponding to this odds ratio value (for example the red curve, for $\phi = 2$).}
        \label{fig:parameterSpaceExamples}
\end{figure}

In treatment/ control test settings, there often exists prior knowledge of a minimal clinically relevant or expected odds ratio $\text{OR}(\theta_a,\theta_b) := (\theta_b/(1 - \theta_b))((1 - \theta_a)/\theta_a)$, i.e. it is known that   $\text{OR}(\theta_a,\theta_b)= \phi$ for some given $\phi$. In that case, one can restrict estimating $\breve\theta_a$ and $\breve\theta_b$ to $\Theta_1(\phi) = \{(\theta_a, \theta_b); \text{OR}(\theta_a, \theta_b) = \phi\}$, possibly improving power and growth-rate of the test \citep{turner2021safe}. Both search spaces are illustrated in Figure \ref{fig:parameterSpaceExamples}.
\paragraph{Combining e-variables from individual strata} We can use the e-variable in \eqref{eq:niceEvar} to calculate e-process values $E^{(m),k}$ 
for each stratum $k$ separately. To be precise, we set $S^k_j$ to the equivalent of (\ref{eq:niceEvar}) if $k= k_j$,
\begin{equation}
\label{eq:newevar}
S^k_j =   
\prod_{i=1}^{n_a}  \frac{p_{\breve\theta_{a,k}|Y^{(j-1)}}(Y_{j,a,i})}{
p_{\breve\theta_{0,k} |Y^{(j-1)}}(Y_{j,a,i}
)} 
\prod_{i=1}^{n_b} 
\frac{p_{\breve\theta_{b,k}|Y^{(j-1)}}(Y_{j,b,i})}{
p_{\breve\theta_{0,k} |Y^{(j-1)}}(Y_{j,b,i})
}, 
\end{equation}
and $S^k_{j} = 1$ otherwise, i.e. if $k_j \neq k$, and $E^{(m),k} := \prod_{j=1}^m S^k_j$ --- note that at each `time $j$', the product e-variable only changes for the $k$ such that $j$-th block was a block of outcomes in stratum $k$. 

We now need to combine the e-processes-per-stratum into a single e-process for \eqref{eq:generalH0} to measure evidence against ${\cal H}_0$ and allowing tests with type-I error probability guarantee on \eqref{eq:generalH0}, the global null hypothesis that the odds ratio of the success probabilities equals $1$ in each stratum. 
There are several ways to do this. 
The first and most straightforward option is to \emph{multiply} the individual e-values across the strata:
\begin{equation}
    \label{eq:simpleMultiplying}
   E^{(m)} = 
   \prod_{j=1}^m S_j^{k_{j}} = \prod_{j=1}^m \prod_{k=1}^K S_j^k.
\end{equation}
To see that $E^{(1)}, E^{(2)}, \ldots$ is an e-process, simply note that each $\smash{S_j^{k_j}}$ is a conditional e-variable (i.e. it satisfies (\ref{eq:evar}) with $S = \smash{S_j^{k_j}}$)  since, given that $S_j$ in (\ref{eq:niceEvar}) is a conditional  e-variable, $\smash{S_j^{k_j}}$  must be an e-variable as well.
When $\theta_{a,k} \approx \theta_{b,k}$ in a few of the strata, this might be a data-inefficient approach, as one would need to collect a lot of extra {evidence} in the strata where the success probabilities are substantially different to counteract the expected small e-values in the other strata.
A second option that possibly better handles these cases is to create a \emph{convex combination}, i.e. a mixture, 
of e-values at each time point $j$ (any convex combination of e-variables is also an e-variable \citep{VovkW21}): 
A simple first option is  to pick some prior distribution on the strata $\pi(k)$, and to use that distribution for calculating the mixture  after each batch comes in\vspace*{-2mm}:
\begin{align}
    \label{eq:simpleAveraging}
&   S_j := \sum_{k = 1}^K \pi(k) S^{k}_{j} \ ; \ E^{(m)}=\prod_{j = 1}^m S_j 
\text{\ so that also} \nonumber   \\
&  \text{$E^{(m)} = \prod_{k=1}^K E^{(m),k}$ 
with $E^{(m),k} = \prod_{j = 1}^m S_j^k$}.
\end{align}
%
Extending the simple averaging above, we could replace the prior $\pi(k)$ in \eqref{eq:simpleAveraging} with a distribution $\pi(k|y^{(j-1)})$ that depends on previous data $y^{(j-1)}$, since, since we assume the data itself in each block are independent, dependency of $\pi$ on past data will not affect guarantee (\ref{eq:evar}).  Such an approach is called the {\em method of mixtures\/} in the anytime-valid testing literature \citep{ramdas2022game}. Thus, any distribution on $[K]$ that depends on the past is allowed here, but an intuitive choice is a {\em pseudo-Bayesian posterior\/}
\begin{equation}
\label{eq:posteriorE}
    \pi(k|y^{(j-1)}) := \frac{\pi(k) (E^{(j-1),k})^{\eta}}{\sum_{k'} \pi(k' ) (E^{(j-1),k'})^{\eta}},
\end{equation}
where by definition, $E^{(0)} = 1$ and we pick $\eta$ beforehand as a \emph{learning rate}:  if we set it to a higher value, we will focus on strata with higher e-values more quickly; with $\eta=1$, (\ref{eq:posteriorE}) becomes similar to a Bayesian posterior. Just as the beta-posterior used to determine $\breve{\theta}_{x,k}$  in (\ref{eq:newevar}) allows us to learn $\theta^*_{x,k}$, this new posterior allows us to learn which strata can help us most to reject the null.
However, even for $\eta=1$ the analogy to Bayes only goes so far --- for example, at each $j$, only the e-variable $S^{k_j}$ for stratum $k_j$ changes; the other $S^k$ `sleep' \citep{koolen2010freezing} and thus $E^{(j-1),k}$ behaves differently from a likelihood. This more general past-determined updating originates in the area of machine learning called {\em prediction with expert advice} where many other such `posterior'-updates have been considered \citep{herbster1998tracking,ErvenGR07,koolen2013universal}. 
\commentout{
With $\eta = 1$, one can also see that \eqref{eq:posteriorE} combined with \eqref{eq:simpleAveraging} result in a more flexible expression through telescoping :
\begin{align}
    E^{(m)} &= \prod_{j = 1}^m \sum_{k = 1}^K \pi(k|y^{(j-1)}) S^{k}_{j} \label{eq:posteriorPerStratum}\\
    & = \prod_{j = 1}^m \sum_{k = 1}^K \frac{\pi(k) E^{(j-1),k}}{\sum_{k' = 1}^K \pi(k' ) E^{(j-1),k'}} S^{k}_{j} \nonumber \\
    &= \prod_{j = 1}^m  \frac{\sum_{k = 1}^K \pi(k) E^{(j-1),k} S^{k}_{j}}{\sum_{k' = 1}^K \pi(k' ) E^{(j-1),k'}} \nonumber \\
    &= \prod_{j = 1}^m  \frac{\sum_{k = 1}^K \pi(k) E^{(j),k}}{\sum_{k' = 1}^K \pi(k' ) E^{(j-1),k'}} \nonumber \\
    &= \frac{\sum_{k = 1}^K \pi(k) E^{(m),k}}{\sum_{k' = 1}^K \pi(k' ) E^{(0),k'}} = \sum_{k = 1}^K \pi(k) E^{(m),k}, \label{eq:easyLearning}
\end{align}
where replacing $E^{(m),k}$ by $E^{(m_k),k}$ would still yield a valid e-variable. In other words, we could still actively learn the best e-variable to put mass on without explicitly calculating a posterior at each time point, and while updating as batches from individual strata come in one-by-one. }
These include the more extreme approach called \emph{switching}. With this approach, we calculate \eqref{eq:simpleAveraging}
with $\pi(k)$ replaced by any distribution we like (the choice is again allowed to depend on $Y^{(j-1)}$) up to and including a particular batch $j^*$. Thereafter, for $j \geq j^*$, we set 
\begin{equation}
\label{eq:switch}
    \pi^*(k | y^{(j)}) = 
    \begin{cases}
    1 & \text{if $k = k^*$ with $k^* =\arg \max_{k} E^{(j^*),k}$} \\
    0 & \text{otherwise}
    \end{cases}
\end{equation}
creating a new E-process $E^{(1)}_{[j^*]}, E^{(2)}_{[j^*]}, \ldots$  such that, for $m \leq j^*$, $E^{(m)}_{[j^*]} = E^{(m)}$ and, for $m > j^*$, 
\begin{equation}
\label{eq:switchE}
E^{(m)}_{[j^*]} = E^{(j^*)} \cdot  \prod_{j=j^*+1}^m E^{(j),k^*}
\end{equation}
$j^*$ could arbitrarily be picked prior to the study, or we could also place a prior on the moment of switching and take a weighted average over \eqref{eq:switchE} for various values of $j^*$ for each $\delta$, thereby obtaining yet another e-process with $j^*$ `integrated out' (see Figure \ref{fig:switchPriorsBoundsMinDifference} in the supplementary material for a more elaborate comparison of switch priors in a simulation experiment for confidence sequences).

\begin{figure}[ht]
\vspace{.3in}
    \centering
    \includegraphics[width=6cm]{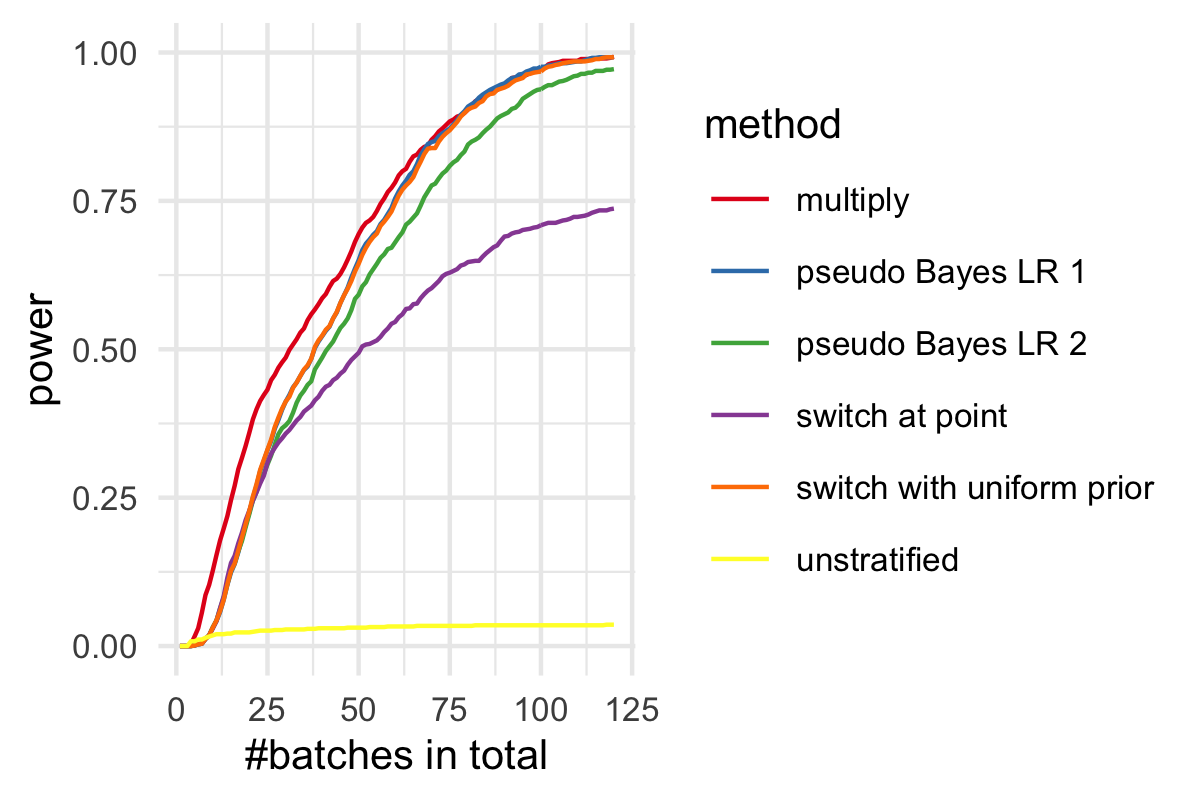}
    \vspace{.3in}
    \caption{Power for rejecting the null at level $\alpha = 0.05$ that the odds ratio in all strata equals 1 estimated with 1000 repeated experiments for various e-variable combination methods. $40$ batches were collected in each of three strata (so maximum sample size was $m=120$) and sampling was stopped as soon as $E^{(m)} \geq \frac{1}{\alpha}$. Real control group success rates were $0.1, 0.2, 0.8$ and real risk differences were $0.05, 0.4, -0.6$. Pseudo-Bayesian approaches were implemented with learning rates (LR) $1$ and $2$. Switch approaches were implemented for switching at point $j^* = 10$, or with a uniform prior on switch times $j = 5$ until $m - 5$.}
    \label{fig:H0combinationspower}
\end{figure}

\begin{figure}[ht]
\vspace{.3in}
    \centering
    \includegraphics[width=7cm]{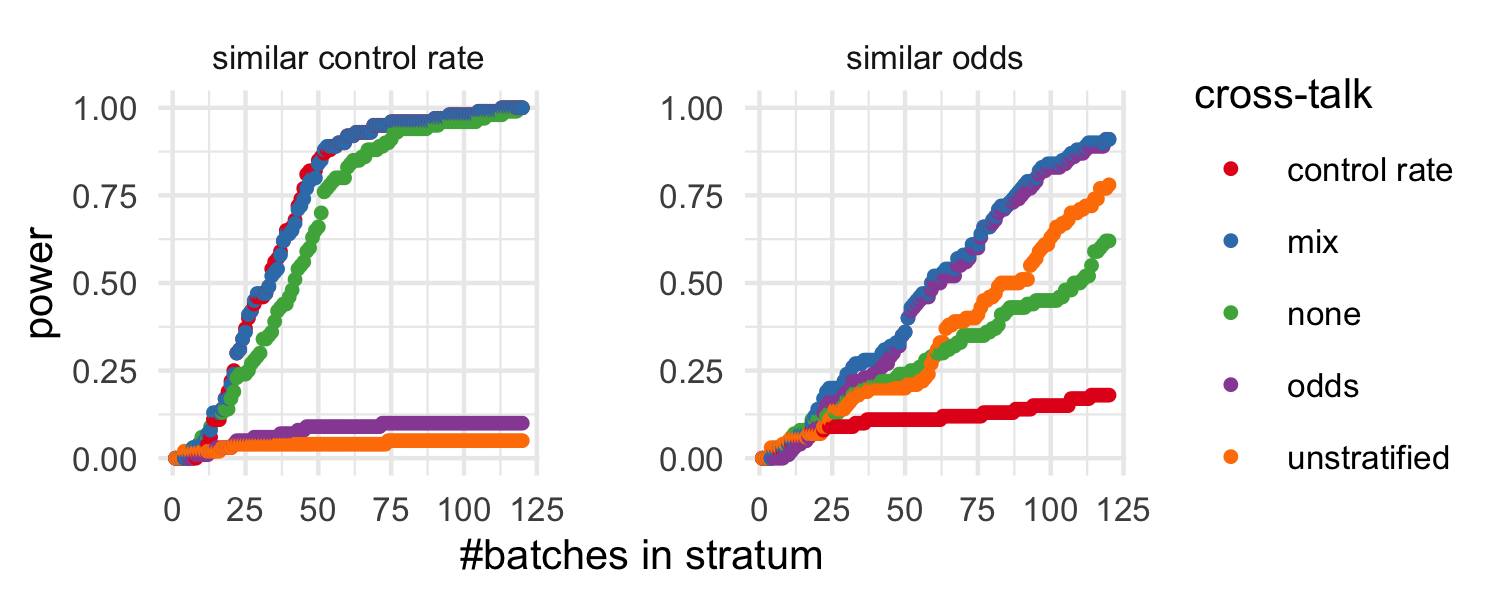}
    \vspace{.3in}
    \caption{Power for rejecting the null hypothesis at level $\alpha = 0.05$ that the odds ratio in all strata equals 1 estimated with 100 repeated experiments for various types of cross-talk. $40$ batches were collected in each stratum and sampling was stopped as soon as $E^{(m)} \geq \frac{1}{\alpha}$. On the left, real control group success rates were $0.49, 0.5$ and $0.51$ in each stratum; risk differences were $-0.09, -0.49, 0.39$. On the right, real odds ratios were $4, 4.01, 2.95$.}
    \label{fig:H0cross-talkpower}
\end{figure}

In Figure \ref{fig:H0combinationspower}, the three different methods for combining e-variables for testing $\Hc_0$ are compared with respect to \emph{power}: the expected probability of rejecting $\Hc_0$ under some fixed data generating distribution. For Figure \ref{fig:H0combinationspower}, data were sampled from a distribution where risk differences and control group rates all differed between strata. It can be observed that all methods that took the stratification into account outperformed the unstratified approach, where just one sequential e-variable was calculated for all strata combined. The three different methods will be re-compared for confidence sequences in Figure~\ref{fig:confIntervalForMinDifference}.
\paragraph{Cross-talk between strata} To further improve power of the hypothesis test, we will allow for {\em cross-talk\/} between strata while estimating $\breve\theta_{a,k}$ and $\breve\theta_{b,k}$ based on data seen so far.
In the current simple setting of testing the global null, `cross-talk' simply amounts to design $S_j$ that grow faster (allowing for faster rejecting of the null) if the alternative satisfies certain {\em constraints}. For example, if one expects treatment effects (say, measured as odds ratios) to be stable (identical)  throughout different strata, but control group recovery rates to vary, one would like cross-talk about the odds ratios between strata. Practically, this means that to arrive at the estimates $\breve\theta_{x,k} \mid Y^{(j-1)}$, we first limit the parameter space to $\Theta_1(\smash{\hat{\phi}^{(j-1)}})$, i.e. all vectors $\theta_{x,k}$ with odds ratio $\hat{\phi}^{(j-1)}$, set to be the maximum likelihood odds ratio based on all previous data in all strata, i.e. calculated by ignoring strata. We then calculate
$\breve\theta_{x,k} \mid Y^{(j-1)}$ as posterior means using beta priors conditioned on the parameters being in $\Theta_1(\hat{\phi}^{(j-1)})$. Similarly, when one expects control group recovery rates to be stable, but the treatment effects to vary because of a possible interaction with stratum characteristics, allowing cross-talk about control group recovery rates might improve power. In practice, we achieve this by using as beta prior parameters for the control group rate $\breve\theta_{a,k}| Y^{(j - 1)}$ the total counts of failures and successes aggregated over all strata (summed with some initial prior parameters to ensure stable estimates at time point $j = 1$; we set initial prior values $0.18$ for both the fail and success rate based on a suggestion by \citep{turner2021safe}). In the odds-ratio cross-talk scenario, we effectively constrain the parameters of the alternative  $\breve\theta_{x,k} \mid Y^{(j-1)}$ at each $j$ to share the same odds-ratio; in the control-group cross-talk scenario, we constrain these parameters to share the same $\theta_{a}$, i.e. $\breve\theta_{a,k}| Y^{(j - 1)}= \breve\theta_{a,k'}| Y^{(j - 1)}$
for each $k,k'$. Would one be unsure whether cross-talk would improve power at all, and if so, whether one should cross-talk on the odds ratios or the cross ratios, one could put prior mass $1/3$ on each of the corresponding three e-values, say $E^{(m)}_{\rho}$ for $\rho \in \{ \textsc{none},\textsc{odds}, \textsc{control rate} \}$, where $\textsc{none}$ stands for the standard e-variable without cross-talk.  One could then, for each block $j$, use a mixture e-variable, where the three e-values are mixed as in \eqref{eq:posteriorE} with $\eta=1$, $k$ replaced by $\rho$ and `$E^{(j-1),k}$' replaced by '$E^{(j-1)}_{\rho}$' giving a new `\textsc{mix}' e-process. All four cross-talk scenarios are explored in simulations in Figure \ref{fig:H0cross-talkpower}, where data were generated from strata with similar control group success rates, but different risk differences, and different control group success rates, but similar odds ratios showing that allowing for cross-talk on control rate or odds ratio improves power in the respective scenarios. The cross-talk mixture performs comparably to the optimal cross-talk options in both cases. 
Cross-talk can be expected to improve power even if, in truth, under the alternative, the odds-ratio resp. control-group rate is just similar, but not exactly the same under all groups; and the confidence sequences of the next section remain valid (but will get wider) even if the odds-ratios resp. control-group rates happen to be completely different. Thus, the method described here cannot really be viewed as a constraint on the model, and we chose to call it  {\em cross-talk\/} instead: data in one stratum informs, `talks to' estimates for other strata.

\begin{figure*}[ht]
\vspace{.3in}
     \centering
     \begin{subfigure}[b]{4.5cm}
         \centering
         \includegraphics[width=\textwidth]{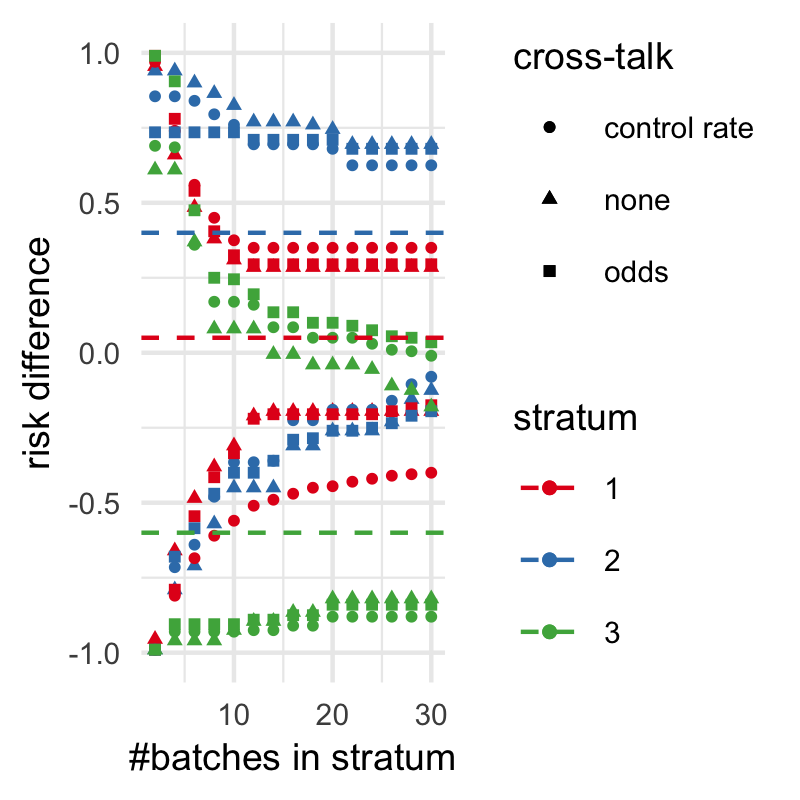}
         \caption{all different}
     \end{subfigure}
     \hfill
     \begin{subfigure}[b]{4.5cm}
         \centering
         \includegraphics[width=\textwidth]{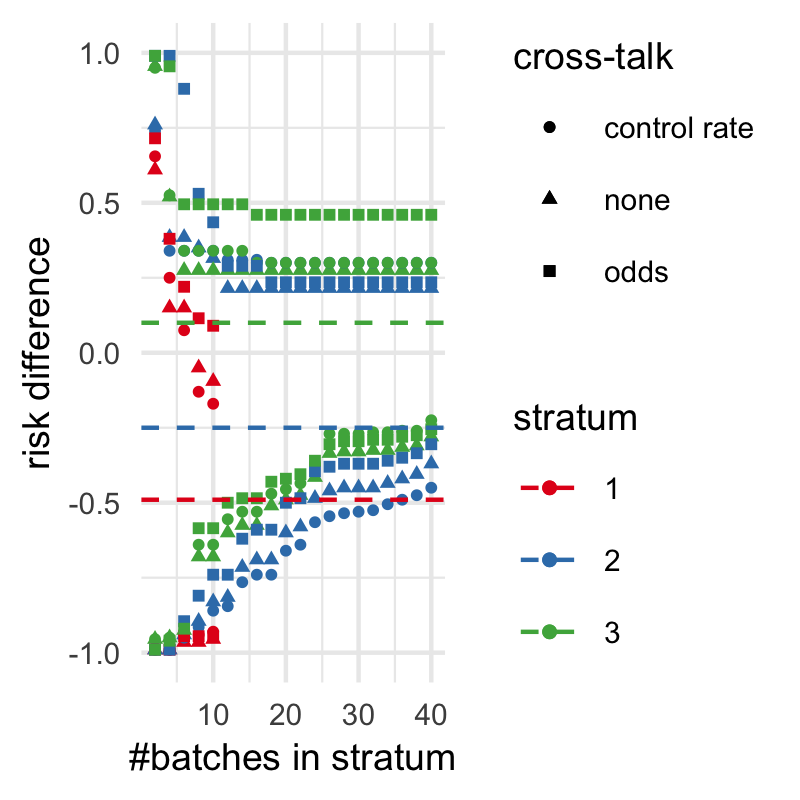}
         \caption{same control group rate}
     \end{subfigure}
     \hfill
     \begin{subfigure}[b]{4.5cm}
         \centering
         \includegraphics[width=\textwidth]{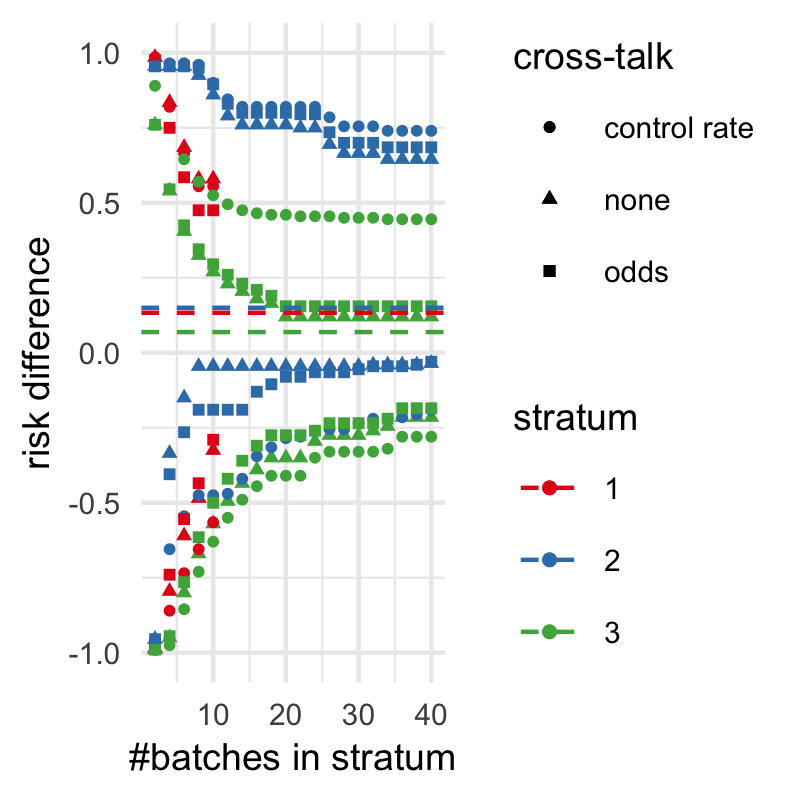}
         \caption{same OR}
     \end{subfigure}
     \begin{subfigure}[b]{4.5cm}
         \centering
         \includegraphics[width=\textwidth]{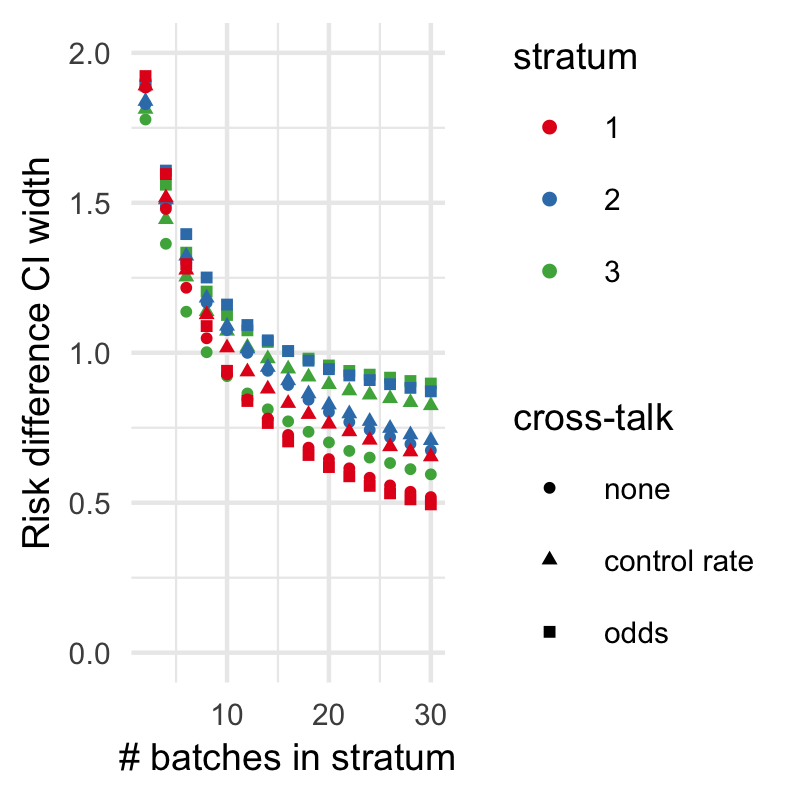}
         \caption{all different}
     \end{subfigure}
     \hfill
     \begin{subfigure}[b]{4.5cm}
         \centering
         \includegraphics[width=\textwidth]{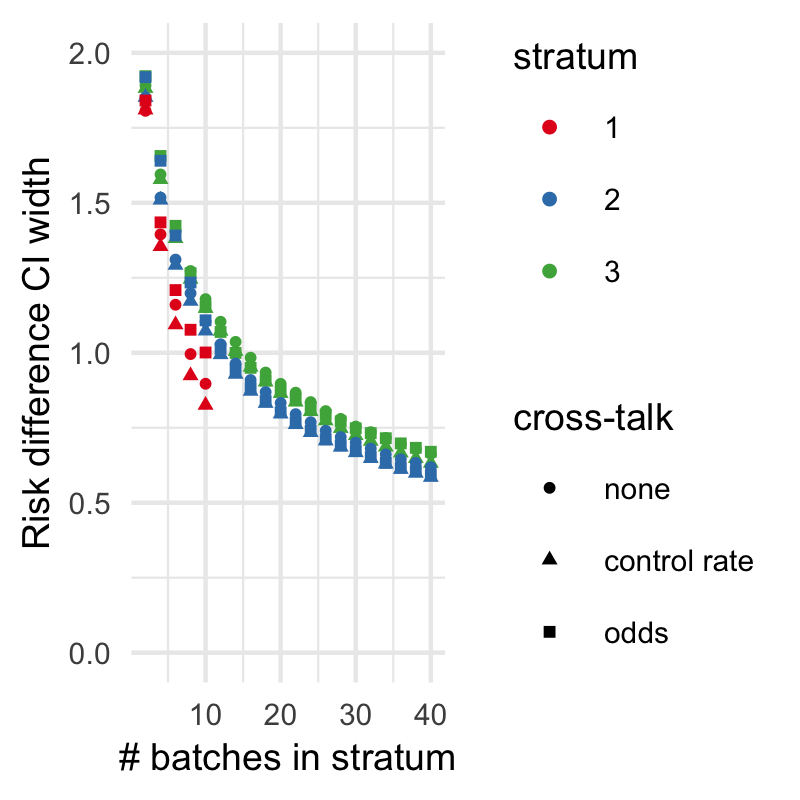}
         \caption{same control group rate}
     \end{subfigure}
     \hfill
     \begin{subfigure}[b]{4.5cm}
         \centering
         \includegraphics[width=\textwidth]{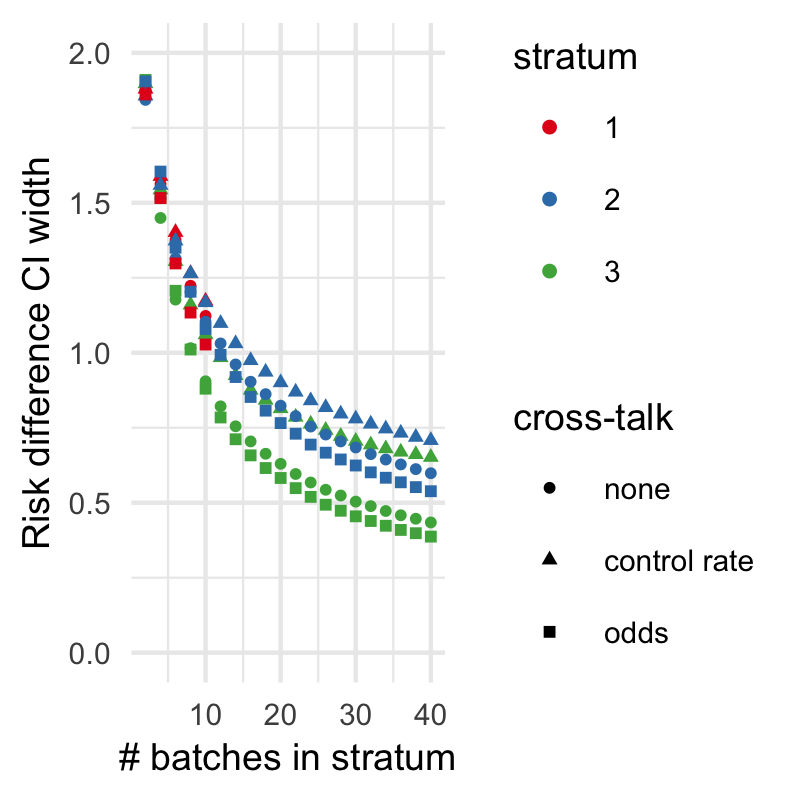}
         \caption{same OR}
     \end{subfigure}
     \vspace{.3in}
        \caption{Examples of $95$\% stratified confidence intervals ((a), (b) and (c)) and mean confidence interval widths estimated over $100$ runs ((d), (e) and (f)) with different types of cross-talk. In (a), (b) and (c) the true risk difference of the data generating distribution in each stratum is indicated by a dashed line. For (a) and (d), the data were generated by distributions with different control group success rates ($0.1$, $0.2$ and $0.8$) and risk differences ($0.05$, $0.4$ and $-0.6$) in each stratum. For (b) and (e), strata sizes were unbalanced: as can be seen for stratum 1, the red points, data collection stopped after $10$ batches. Control group success rates were all $0.5$ and risk differences were different ($-0.49$, $-0.25$ and $0.1$). For (c) and (f), strata sizes were unbalanced as well, and now odds ratios were the same in each stratum ($2$), but control group rates differed again ($0.2$, $0.25$ and $0.85$). }
        \label{fig:crossTalkExamples}
\end{figure*}
\paragraph{A GRO-Sanity Check}
While the simulations above and below show encouraging empirical results regarding the power of our methods, it is still useful to have some theoretical assurance that, no matter the `true' alternative generating the data, all methods we consider produce e-values that grow fast (i.e. achieve good power) under this alternative. We now provide a simple theorem to this end. As usual in the e-value and safe-testing literature, and for reasons explained by \cite{grunwald2019safe},  we concentrate on GRO (\ref{eq:grodef}) rather than power.
\begin{thm}\label{thm:GROconfseq} Suppose that we observe $m = m_1 + \ldots + m_K$ blocks, with $m_k$ blocks lying in stratum $k$, each such block sampled independently from $P_{\theta^*_{a,k},\theta^*_{b,k}}$.
Then, with ${\bf E}$ denoting expectation under this distribution, the e-process $E^{(m)}$ defined by multiplication as in (\ref{eq:simpleMultiplying}) and  the {\sc mix\/} e-process $E^{(m)}$ as  above with constituent e-processes defined multiplicatively as in (\ref{eq:simpleMultiplying}) 
both achieve: 
\begin{equation}\label{eq:mkgro}
 \sum_{k=1}^K  m_k \textsc{gro}(\theta^*_{a,k},\theta^*_{b,k}) =
    {\bf E} \left[  \log E^{(m)}  \right] +  O(\log m). 
\end{equation}
\end{thm}
To interpret the result, note that, if an oracle were to supply us with $\theta^*_a = (\theta^*_{a,1}, \ldots, \theta^*_{a,k}),\theta^*_b = (\theta^*_{b,1}, \ldots, \theta^*_{b,k})$ i.e. if we were told  `if the alternative were true, then its parameters would be $P_{\theta^*_a,\theta^*_b}$', then we could use the GRO (growth optimal e-variable) which, conditional on observing a block in stratum $k$, would obtain the optimal, largest possible expected growth $\textsc{gro}(\theta^*_{a,k},\theta^*_{b,k})$. Since we assume data to be independent, the best growth we could obtain with such an oracle is given by the left-hand side of (\ref{eq:mkgro}). The theorem expresses that the price for learning (via Bayes predictive distributions $\breve{\theta}_{x,k}$ based on beta-priors) rather than knowing $\theta^*_a,\theta^*_b$ is modest, namely logarithmic in $m$  whereas the growth itself is linear in $m$; this is the standard situation for parametric settings, described in detail by \cite{grunwald2019safe}. We may expect the constant hidden in the $O(\log m)$ to become substantially smaller if the preconditions for effective cross-talk hold as described above, e.g. odds ratios or group recovery rates are identical or similar across strata; but determining this constant precisely across cases, as well as extending the analysis to pseudo-Bayesian and switch e-processes, is complicated and will be left for future research. The proof of this theorem can be found in the appendix.  
\section{EXTENSION TO CONFIDENCE SEQUENCES}
\label{sec:confidence}
\cite{turner2022confseq} showed that \eqref{eq:niceEvar} in the $2 \times 2$-table (single stratum) can be generalized, to test null hypotheses $\mathcal{H}_0 := \{ P_{(\theta_a, \theta_b)}; (\theta_a, \theta_b) \in \Theta_0 \}$
beyond `$\theta_a = \theta_b$':
\begin{align}\label{eq:woensdag}
S_{j,[\Theta_0]}
= 
\prod_{i=1}^{n_a} \frac{p_{\breve\theta_a|Y^{(j-1)}}(Y_{j,a,i})}{
p_{\breve\theta_a^{\circ} |Y^{(j-1)}}(Y_{j,a,i})}
\prod_{i=1}^{n_b} 
\frac{p_{\breve\theta_b|Y^{(j-1)}}(Y_{j,b, i})}{
p_{\breve\theta^{\circ}_b |Y^{(j-1)}}(Y_{j, b,i})
} 
\end{align}
is an e-variable, as long as $\Theta_0 \subset [0,1]^2$ is convex and closed.  
Here
$(\breve\theta^{\circ}_a \mid Y^{(j-1)}, \breve\theta^{\circ}_b \mid Y^{(j-1)})$ is defined to minimize KL divergence, i.e.  is the pair $(\theta_a,\theta_b) \in \Theta_0$ that minimizes, over $\Theta_0$, \\
$
D(P_{\breve\theta_a \mid Y^{(j-1)},\breve\theta_b \mid Y^{(j-1)}}(Y^{n_a}_a,Y^{n_b}_b) \|  P_{\theta_a,\theta_b}(Y^{n_a}_a,Y^{n_b}_b))$.
(\ref{eq:niceEvar}) is a special case since with $\Theta_0 = \{(\theta,\theta): \theta \in [0,1]\}$, this KL divergence is minimized by $(\breve\theta^{\circ}_0,\breve\theta^{\circ}_0)$ with $\breve\theta^{\circ}$ as defined underneath (\ref{eq:niceEvar}). 
Again, $\breve\theta_a$  and $\breve\theta_b$ are estimated based on past data $Y^{(j-1)}$ as in \eqref{eq:niceEvar}. 
Based on (\ref{eq:woensdag}) one can construct an {\em exact\/} (nonasymptotic) confidence sequence (CS)
\begin{equation}
    \label{eq:avci}
\text{\sc CS}_{\alpha,(m)} = \left\{ \delta : 
E^{(m)}_{[\Theta_0(\delta)]} \leq \frac{1}{\alpha} \right\},
\end{equation}
with $\Theta_0(\delta) \subset [0,1]^2$ a null hypothesis determined by a divergence measure. By construction, such a confidence sequence is  {\em always-valid\/} \citep{ramdas2022game} in the sense that for any $\delta$, any $\theta \in \Theta_0(\delta)$, the $P_{\theta}$-probability that there will {\em ever\/} be an $m$ such that $\delta \not \in \text{\sc CS}_{\alpha,(m)}$ is at most $\alpha$. This means that we can take the \emph{running intersection} of the confidence sequence while retaining coverage, which will be used throughout the simulation experiments in this paper.
In this paper, we are going to construct confidence sequences for risk differences as examples, where we are going to test hypotheses of the form $\Theta_0(\delta) := \left\{(\theta_a,\theta_b) \in [0,1]^2: \theta_b - \theta_a = \delta \right \}$ --- below we extend this to the case that differentiates in terms of the strata. Still, everything could also easily be adapted to construct confidence intervals for other divergence measures, such as odds and risk ratios \citep{turner2022confseq}. 
    
\subsection{One CS per stratum} If we expect the effect size values to differ between the strata, one could decide to report a separate confidence sequence for each stratum using (\ref{eq:avci}) above. To reach a better estimate sooner, we could however still allow cross-talk on control group success rates or odds ratios between subpopulations, as described in section 2 above. In this setup, we would end up with a \emph{collection} of $k$ confidence sequences:

\begin{equation}
\text{\sc CS}^k_{\alpha,(m)} = \left\{ \delta : 
 E^{(m), k}_{[\Theta_0(\delta)]} \leq \frac{1}{\alpha} \right\},
\end{equation}
with $\breve\theta_a$ and $\breve\theta_b$ in $E^{(m), k}$ estimated based on data seen up to time $m$ and $E^{(m),k}$ defined as in (\ref{eq:simpleAveraging}) with $S^k_j$ replaced by $S^k_{j,[\Theta_0]}$ as in (\ref{eq:woensdag}), calculated for stratum $k$.   
Illustrations of confidence intervals over time with the three options for cross-talk are depicted in Figure \ref{fig:crossTalkExamples}. As can be observed there, not allowing cross-talk gives the best results when the true data generating distributions in the strata have different control group success rates and odds ratios (see the circle-shaped points in Figure \ref{fig:crossTalkExamples}d, especially in the third stratum, where the effect size has a different sign). However, when control group rates or odds ratios are similar across strata, allowing cross-talk improves results. See for example Figure \ref{fig:crossTalkExamples}e, where interval width decreases much faster in the smaller stratum 1 while allowing cross-talk about the control group rate. Similar experiments for comparing confidence sequences with and without the mixture of cross-talk methods can be found in the supplementary material, Figure \ref{fig:crossTalkExamplesMix}.  
    
\subsection{CS for the minimum or maximum} In some scenarios, for example when we do not have the means to collect a large data sample, or when data is very unbalanced in one or more strata, it could be more informative to create one CS for the minimum or maximum effect size value over all strata. To achieve this, we introduce two new forms of null hypotheses and corresponding e-variables that will subsequently be inverted to create two one-sided confidence sequences, for lower and upper bounds on the minimum or maximum. 

\begin{figure}[ht]
\vspace{.3in}
    \centering
    \begin{subfigure}[b]{5cm}
    \centering
    \includegraphics[width = \textwidth]{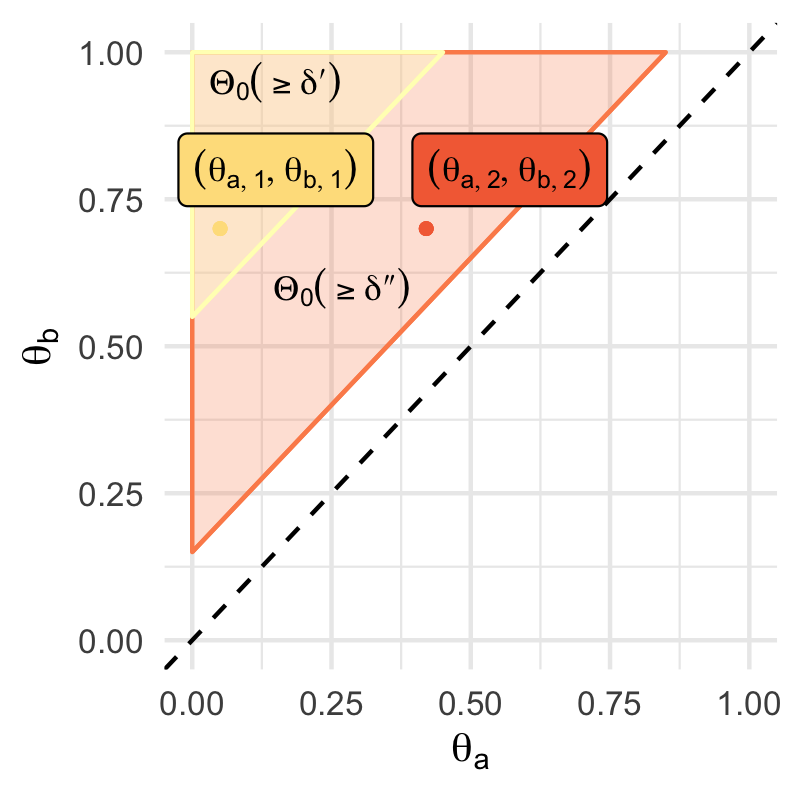}
    \caption{Examples of parameter spaces for $\mathcal{H}_{0, \delta}: \forall k: \theta_k \in \Theta_0(\geq \delta)$.}
    \label{fig:greaterThanParamSpace}
    \end{subfigure}
    \hspace{1cm}
    \begin{subfigure}[b]{5cm}
    \centering
    \includegraphics[width = \textwidth]{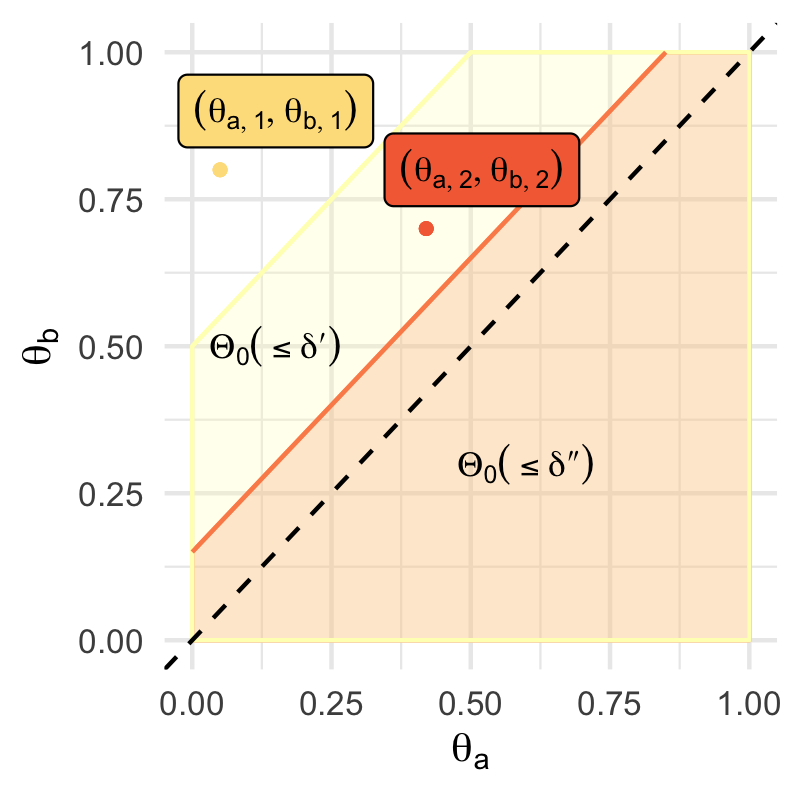}
    \caption{Examples of parameter spaces for $\mathcal{H}_{0, \delta}: \exists k: \theta_k \in \Theta_0(\leq \delta)$.}
    \label{fig:smallerThanParamSpace}
    \end{subfigure}
    \vspace{.3in}
    \caption{Parameter space examples for hypotheses tested to construct upper and lower bounds on minima and maxima of effect size values}
\end{figure}

\paragraph{One-sided CS: upper bound} We will first illustrate how to estimate an upper bound on some minimal effect size value over strata\footnote{Analogously, with this method a lower bound on some maximal effect size value can be estimated by reversing all signs.}. To this end, we consider a null hypothesis of the form $\mathcal{H}_{0, \delta}: \forall k: \theta_k \in \Theta_0(\geq \delta)$ (i.e. for risk difference effect size, $\Theta_0(\geq \delta)= \{(\theta_a,\theta_b) \in [0,1]^2 : \theta_b -\theta_a \geq \delta\}$) and aim to design e-variables to test it. E.g. in the example depicted in  Figure \ref{fig:greaterThanParamSpace}, we aim to design an e-variable that will reject $\mathcal{H}_{0, \delta''}$ at any batch $j$ with probability less than $\alpha$ (i.e., that offers type-I error guarantee), when the data in the strata are in reality generated by $(\theta_{a,1}, \theta_{b,1})$ and $(\theta_{a,2}, \theta_{b,2})$. We do eventually want to reject $\mathcal{H}_{0, \delta'}$ as $\delta((\theta_{a,2}, \theta_{b,2})) < \delta'$. As we collect more and more data, we can reject null hypotheses corresponding to values of $\delta'$ for which $\delta' - \delta((\theta_{a,2}, \theta_{b,2}))$ gets closer and closer to $0$.

Let us denote the e-process consisting of the e-variables for testing $\theta_k \in \Theta_0(\geq \delta)$ in each stratum combined, using any of the methods described above in Section 2, as $E^{*(m)}_{\delta}$. 
The one-sided confidence interval for the minimum effect can be defined as: 
\begin{equation}
\label{eq:upperBoundMinimum}
    \text{CS}^+_{\alpha,(m)} := \left[-1, \min \left\{\delta: E^{*(m)}_{\delta} \geq \frac{1}{\alpha}\right\}\right].
\end{equation}
All possible approaches for combining e-variables from separate strata, as described in Section 2 above, to find an upper bound for the minimal effect size value are compared in the confidence intervals in the paragraph below.

\begin{figure}[htbp]
\vspace{.3in}
    \centering
    \includegraphics[width = 6cm]{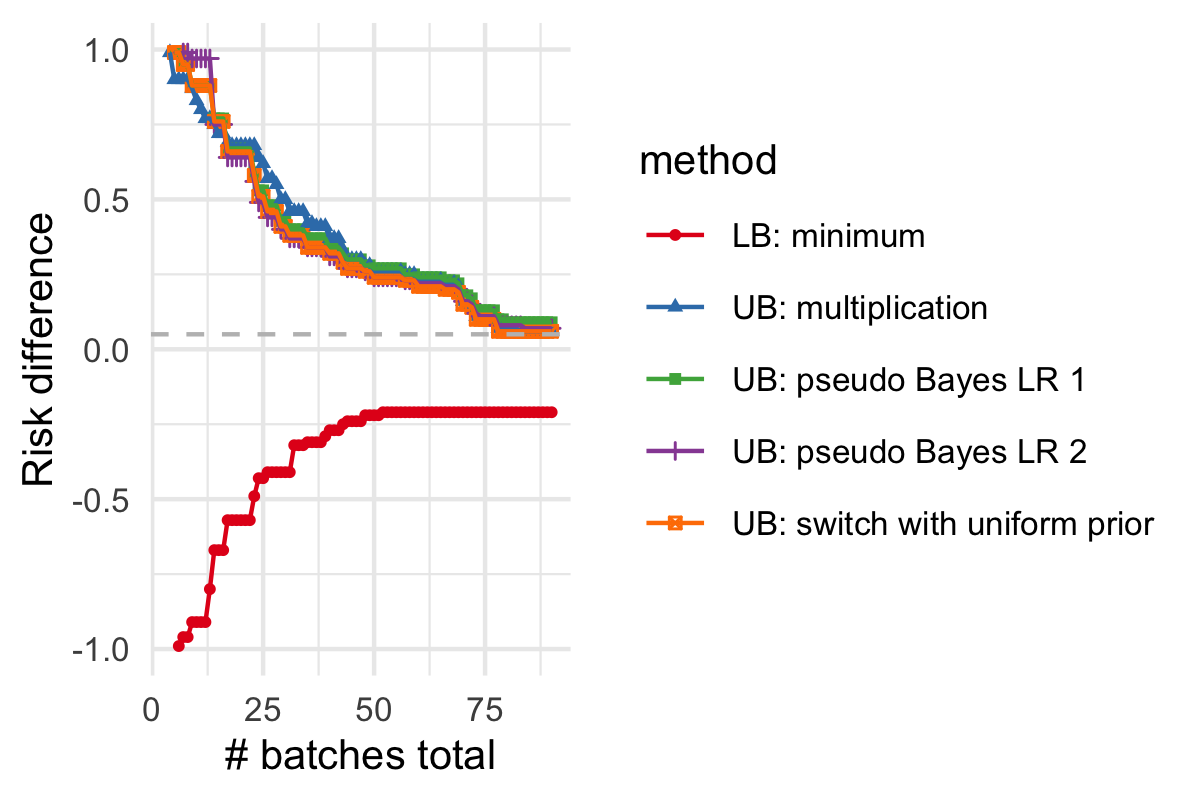}
    \vspace{.3in}
    \caption{Example of confidence sequences for the lower- (LB) and upper (UB) bounds of the minimum effect. $30$ observations were made in each stratum, and the real differences were $0.5$, $0.4$ and $0.05$. With the switch method, a uniform prior ranging from $m_{\text{switch}} = 5$ until $30$ was applied. With the pseudo-Bayesian approach, the learning rate $\eta$ was set to $1$ and $2$. $\alpha$ was set to $0.05$. }
    \label{fig:confIntervalForMinDifference}
\end{figure}

\paragraph{One-sided CS: lower bound} We now also aim to estimate a lower bound for the minimal effect size value (or, analogously, an upper bound for the maximal effect size value). To achieve this, we now consider a null hypothesis of the form $\mathcal{H}_{0, \delta}: \exists k: \theta_k \in \Theta_0(\leq \delta)$. Looking at Figure \ref{fig:smallerThanParamSpace} as an example, where data are generated by $(\theta_{a,1}, \theta_{b,1})$ and $(\theta_{a,2}, \theta_{b,2}),$ we aim to design an e-variable that will reject $\mathcal{H}_{0,\delta'}$ at any batch $j$ with probability less than $\alpha$ (i.e., we again want type-I error guarantee if  $\mathcal{H}_{0,\delta'}$ is true), as $\delta((\theta_{a,2}, \theta_{b,2})) < \delta'$. We do want to reject as quickly as possible $\mathcal{H}_{0,\delta''}$, as $\forall k,\delta(\theta^{(k)}) > \delta''$. As we collect more data, we can reject null hypotheses with values of $\delta''$ for which $\delta((\theta_{a,2}, \theta_{b,2})) - \delta''$ gets closer and closer to 0. 

To build our one-sided confidence interval $\text{CS}^-_{\alpha,(m)}$, we again want to construct a compound e-variable $E^{*(m)}_{\delta}$ testing the null hypothesis corresponding to each value of $\delta$, but now take $\max \{\delta: E^{*(m)}_{\delta} \geq 1/\alpha\}$ as our lower bound.
To test $\mathcal{H}_{0, \delta}$ we will use the \emph{minimum} of $E^{(j),k}_{\Theta_0(\leq \delta)}$ over all $k$, which provides an e-variable for ${\cal H}_{0,\delta}$. To see this, let us assume  $\mathcal{H}_{0, \delta}$ is true an that for some $k^*$, $\theta_{k^*}  \in \Theta_0(\leq \delta)$; the other data generating distributions might or might not come from $\Theta_0(\leq \delta)$. Then:
$   \mathbb{E}(\min_k S^k) \leq \min_k \mathbb{E}(S^k) \leq \mathbb{E}(S^{k^*}) \leq 1. $
\paragraph{Combining into confidence interval}
We now combine the lower bound and upper bound estimation methods established above to build confidence intervals for the minimal effect size value. This can be achieved through taking the intersection of the one-sided confidence sequences introduced above: $\text{CS}_{\alpha,(m)} := \text{CS}^-_{\alpha,(m)} \cap \text{CS}^+_{\alpha,(m)}.$ Results from an experiment where in one of the strata the treatment effect was substantially smaller than in the others are depicted in Figure \ref{fig:confIntervalForMinDifference} (with average interval widths in the supplementary material, \ref{fig:CImindifferencsavwidth}). In early phases of data collection, multiplication gives the quickest convergence, but as more data is collected, the ``sequential learning" methods converge quicker. When risk differences where about the same across all strata, multiplication converged the quickest (see Figure \ref{fig:confIntervalForMinDifferenceEqual} in the Supplementary material).

\subsection{CS for the mean effect}
In addition to estimating the minimum or maximum effect in one of the strata, one might be interested in estimating the mean effect an intervention will have on an entire population, given the existence of subpopulations. For example, one might want to estimate the effect a vaccination will have on the probability of people being contaminated with a disease, taking into account that a certain proportion of the population concerns elderly or immunocompromised citizens. 

Assuming we have a trustworthy estimate of the proportion of subjects belonging to each stratum $k$ in the population of interest, $\pi_{k}$, we aim to estimate the mean risk difference (mean expected effect of the intervention) $\delta^* := \sum_k \pi_k \delta_k$. We can build a confidence sequence for $\delta^*$ by constructing an e-variable for the set of all possible success probability distributions satisfying this $\delta^*$, $\Hc_{0, \delta^*}: \{ P_{\vec{\theta}}; d(\vec{\theta}) = \sum_k \pi_k d((\theta_{a,k}, \theta_{b,k})) = \delta^*\}$. It is not directly clear what an optimal e-variable could look like; one option that offers both the type-I error guarantee with potentially good power is to combine the growth-rate optimal e-variable \eqref{eq:niceEvar} for a specific $\delta_k$ in each stratum with the \emph{universal inference} \citep{wasserman2020universal} method for designing e-processes. Based on this strategy, we look at the set of all vectors $\vec{\delta} := (\delta_1, ..., \delta_K)$ that satisfy $\sum_k \pi_k \delta_k = \delta^*$. For one member of the set, we can calculate the e-variable \emph{based on all batches of data seen up to and including time $m$} according to \eqref{eq:niceEvar}:
\begin{equation*}
    E^{(m)}_{[\vec{\delta}]} = \prod_k E^{(m),k}_{[\Theta_0(\delta_k)]},
\end{equation*}
where $\smash{E^{(m),k}_{[\Theta_0(\delta_k)]}}$ can be calculated using estimates for $\breve\theta_{a,k}$ and $\breve\theta_{b,k}$ as before, only including data seen up to \emph{and not including} batch $m$. The e-variable for $\Hc_{0, \delta^*}$ can then be calculated as \citep{wasserman2020universal}:
$E^{*(m)}_{\delta^*} = \min_{\vec{\delta}}  \smash{E^{(m)}_{[\vec{\delta}]}}$,
and the corresponding confidence sequence can be constructed as before, analogously to \eqref{eq:upperBoundMinimum}. 
\begin{figure}[htbp]
\vspace{.3in}
    \centering
    \includegraphics[width = 5cm]{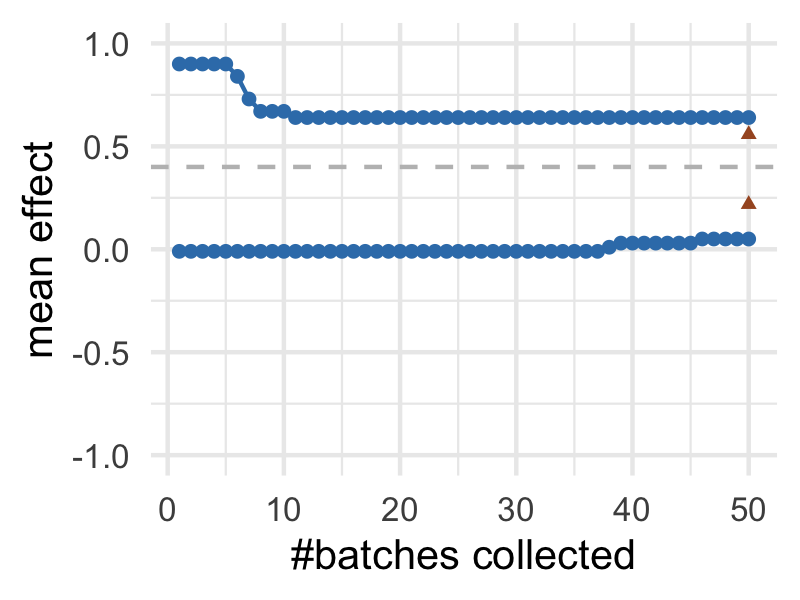}
    \vspace{.3in}
    \caption{Simulated example of $95$\% confidence sequences for the mean effect across subpopulations. $25$ observations were made in each stratum, and the real risk difference of $0.4$ was homogeneous across subpopulations. The confidence sequence for the mean effect is plotted alongside the Miettinen-Nuninen confidence interval, a fixed-n confidence interval method, at batch number $50$ (the purple triangles). In the supplementary materials, figure \ref{meanvsminCI}, the mean effect CS is further illustrated for heterogeneous risk differences in strata.}
    \label{fig:meanEffect}
\end{figure}
\paragraph{Comparison to fixed-n CI for Mantel-Haenszel risk difference} 
Much of the research into estimating stratified risk differences with coverage guarantee has considered Mantel-Haenszel risk differences, where risk differences or odds ratios are \emph{homogeneous} across strata but control group rates can vary (see for example \citep{qiu2019construction}), with fixed-n designs. This is a  strong assumption, and we do not make it ourselves; but we {\em can\/} use cross-talk on the risk difference to  tailor our confidence sequences so that they adapt (get narrow) if the risk difference is indeed homogeneous. One recent fixed-n approach for this setting was described and implemented by \cite{klingenberg2014new}. In Figure \ref{fig:meanEffect}, our confidence sequence for the mean effect is compared to the Miettinen-Nuninen (MN) confidence interval from \cite{klingenberg2014new} at fixed time $50$ in a setting where risk differences were homogeneous. The 
MN-interval is slightly narrower, but because we are allowed to continuously monitor the confidence interval while retaining coverage with the confidence sequence, we can exclude 0 from the CS considerably earlier than with the fixed-n method --- which is remarkable because unlike the MN fixed-n confidence interval, our anytime-valid confidence sequences are also valid if in fact risk differences are not homogeneous. 

\section{CONCLUSION AND FUTURE WORK}
We have introduced a new method for global null hypothesis testing and constructing exact anytime-valid confidence sequences in stratified count data. Our method is complementary to previously proposed methods for similar settings as we need no stochastic assumptions about the arrival times of the subgroups or strata, and no Model-X assumptions. We have shown that our tests and estimates are efficient in terms of power, and that precise effect size estimations can be reached with less strong model assumptions compared to pre-existing fixed-n methods, while retaining coverage guarantees and allowing sequential decision making. We have also shown that we can improve the traditional model of global null testing in the CMH-setting through incorporating ideas from machine-learning: allowing for cross-talk between strata, and incorporating pseudo-Bayesian learning and switching between strata for learning compound effect measures. 

Our work extends that of \cite{turner2021safe} and \cite{turner2022confseq} to incorporate strata for count data. Their methods, however, are generally implementable for any convex null hypothesis, and future work should explore if they also can feasibly be extended to stratified sequential effect estimation for continuous outcome variables.

\subsection{Acknowledgements}
This work is part of the Enabling Personalized Interventions (EPI) project, which is supported by the Dutch Research Council (NWO) in the Commit2Data –Data2Person program, contract 628.011.028.

\bibliography{references}

\newpage

\setcounter{section}{0}
\setcounter{figure}{0}
\appendix
\onecolumn

\renewcommand{\thesection}{S\arabic{section}}
\renewcommand\thefigure{S\arabic{figure}}
\section{PROOFS}
\begin{proof} (of theorem \ref{thm:GROconfseq}).
First consider the basic case with $E^{(m)}$ as in (\ref{eq:simpleMultiplying}).
As we show below, we have, with ${\bf E} \equiv {\bf E}_{P_{\theta^*}}$,
\begin{align}\label{eq:derivation}
&     {\bf E} \left[ \log E^{(m)} \right] = 
 {\bf E} \left[ \sum_{j=1}^m \log S_j \right] = 
{\bf E} \left[ 
     \sum_{j=1..m} \sum_{x \in \{a,b\}} \sum_{i=1..n_x} \log 
     \frac{p_{\breve\theta_{x,k_j}| Y^{(j-1)}} (Y_{j,x,i})}{
     p_{\breve\theta_{0,k_j}| Y^{(j-1)}} (Y_{j,x,i})}  \right] \geq \nonumber \\
     &     {\bf E} \left[ 
     \sum_{j=1..m} \sum_{x \in \{a,b\}} \sum_{i=1..n_x} \log 
     \frac{p_{\breve\theta_{x,k_j}| Y^{(j-1)}} (Y_{j,x,i})}{
     p_{\tilde\theta_{0,k_j}} (Y_{j,x,i})}  \right] \geq  
     {\bf E} \left[ 
     \sum_{\substack{j=1..m \\ x \in \{a,b\} \\ i=1..n_x}} \log 
     \frac{p_{\theta^*_{x,k_j}} (Y_{j,x,i})}{
     p_{\tilde\theta_{0,k_j}} (Y_{j,x,i})}  - \sum_{\substack{k=1..K \\ x \in \{a,b\}}} 
     \log \left( n_x  m_k \right)\right] + O(1) =  \nonumber \\
     &  \sum_{k=1..K} m_k \cdot D( P_{\theta^*_{a,k},\theta^*_{b,k}}
     \| P_{\tilde\theta_{0,k},\tilde\theta_{0,k}}
     ))  + O(\log m)
\end{align}
where  we use notation $D(P_{\theta^*_{a},\theta^*_{b}}
     \| P_{\theta_{0},\theta_{0}}     )$ as in (\ref{eq:kl}); and $\tilde\theta_{0,k}$ is defined as $\arg \min_{\theta \in [0,1]} D(P_{\theta^*_{a,k},\theta^*_{b,k}} \| P_{\theta,\theta}) $  which by the same calculation as the one leading up to (\ref{eq:kl}, is given by $
\tilde\theta_{0,k} = (n_a/n) \theta^*_{a,k} + (n_b/n) \theta^*_{b,k}$, and   
$m_k$ denotes the number of times that an instance of block $k$ was observed in the first $m$ blocks, and
we remind the reader that $+O(\log m)$ may also indicate a negative difference of order $\log m$.  
(\ref{eq:derivation}) immediately implies the result, using (\ref{eq:klgro}). 

The first two equalities in (\ref{eq:derivation}) are immediate. The first 
inequality follows because $P_{\tilde\theta_{0,k_j},\tilde\theta_{0,k_j}}$ minimizes KL divergence to $P_{\theta^*_{a,k_j},\theta^*_{b,k_j}}$ among all $\theta \in [0,1]$, within each block $j$. The final equality  follows by independence and basic calculus. 
It remains to show the second inequality. This one follows because we use a prior $W(\theta_{a,k},\theta_{b,k}$ under which $\theta_a$ and $\theta_b$ are independently  beta distributed with  strictly positive densities on $(0,1)$.  We can then use a standard Laplace approximation of the Bayesian marginal likelihood to obtain, for each fixed $k \in \{1, \ldots, K\}$, where the expectation ${\bf E}$ is over 
$Y'_{(1)}, \ldots, Y' _{(m')} \sim P_{\theta^*_{a,k},\theta^*_{b,k}}$:
\begin{align*}
& {\bf E} \left[ - \log \prod_{j=1}^{m'}  
\prod_{x \in \{a,b\}} \prod_{i=1}^{n_x} {p_{\breve\theta_{x,k}| Y^{(j-1)}} (Y_{j,x, i})}
\right]  = 
{\bf E} \left[ - \log \left( \int 
\prod_{j=1}^{m'} \prod_{x \in \{a,b\}} \prod_{i=1}^{n_x} {p_{\theta_{x,k}} (Y_{j,x, i})} \right) d W(\theta_{a,k},\theta_{b,k}) 
\right]  \\ 
& \leq {\bf E} \left[ \sum_{j=1}^{m'} - \log p_{\theta^*_{a,k},\theta^*_{b,k}}(Y_{(j)}) \right]
+  \log (n_a +n_b) m' + O(1). 
\end{align*}
Here the equality is standard telescoping of the Bayesian marginal likelihood, and the inequality is the Laplace approximation, i.e. the  same calculation as the one leading up to the $(d/2) \log n$ BIC approximation of Bayesian marginal likelihood for a $d$-parameter exponential family; here $d=2$ since we have two free parameters, $\theta^*_{a,k}$ and $\theta^*_{b,k}$; see \cite[Chapter 8]{Grunwald07} for proof and detailed explanation).

This shows the result for the basic case that $E^{(m)}$ is arrived at by multiplication, (\ref{eq:simpleMultiplying}). The case for $E^{(m)}_{\textsc{mix}}$ follows similarly by noting that, by construction,
$E^{(m)}_{\textsc{mix}} \geq E^{(m)}_{\textsc{none}}/3$, where $E^{(m)}_{\textsc{none}}$ denotes the standard e-process with multiplication and without cross-talk, for which we have already (just) shown the result.
\commentout{
The case for the pseudo-Bayesian e-variable with $\eta=1$  follows by noting that, by a standard telescoping argument, the first equality in (\ref{eq:derivation}) can then be replaced by
$$
{\bf E} \left[ \log E^{(m)} \right] 
= {\bf E} \left[ \log \sum_{k=1}^K \pi(k) \prod_{j=1}^m S_j^k \right]
\geq  
 {\bf E} \left[ \max_k \left(\log\left(\prod_{j=1}^m S_j^k\right) + \log \pi(k) \right) \right]
$$
and the result follows by re-tracing the same steps as in (\ref{eq:derivation}) starting from the second equality of that derivation, putting indicator functions in front of the logarithm that are $0$ except when $k_j$ is equal to the $k$ achieving the maximum above. The case with the switch distribution with uniform prior over $j^*$ is handled similarly. }
\end{proof}
\newpage
\section{ADDITIONAL EXPERIMENTS}
\begin{figure}[htbp]
\vspace{.3in}
     \centering
     \begin{subfigure}[b]{4.5cm}
         \centering
         \includegraphics[width=\textwidth]{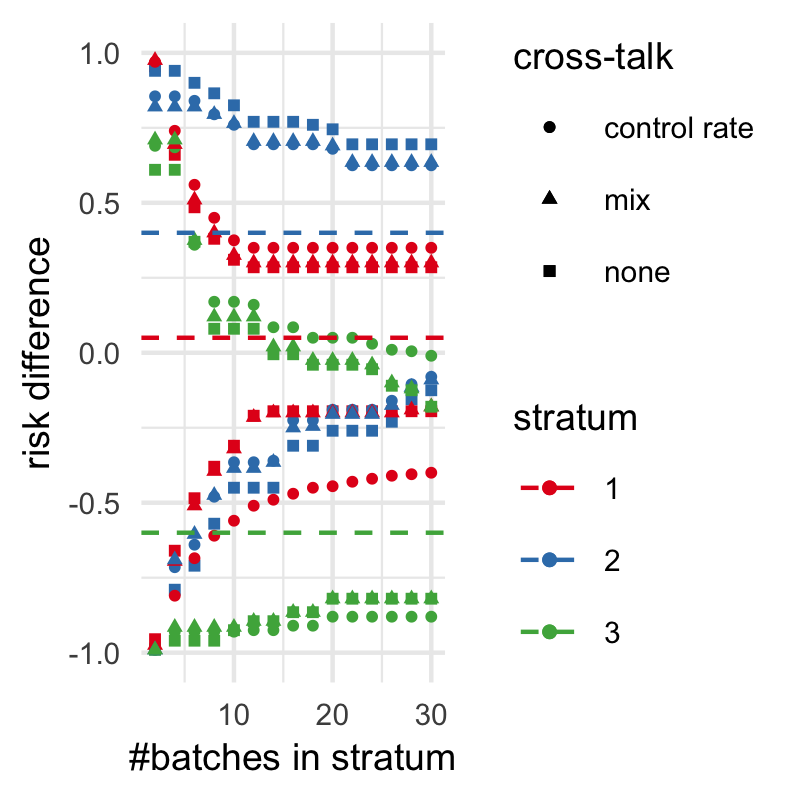}
         \caption{all different}
     \end{subfigure}
     \hfill
     \begin{subfigure}[b]{4.5cm}
         \centering
         \includegraphics[width=\textwidth]{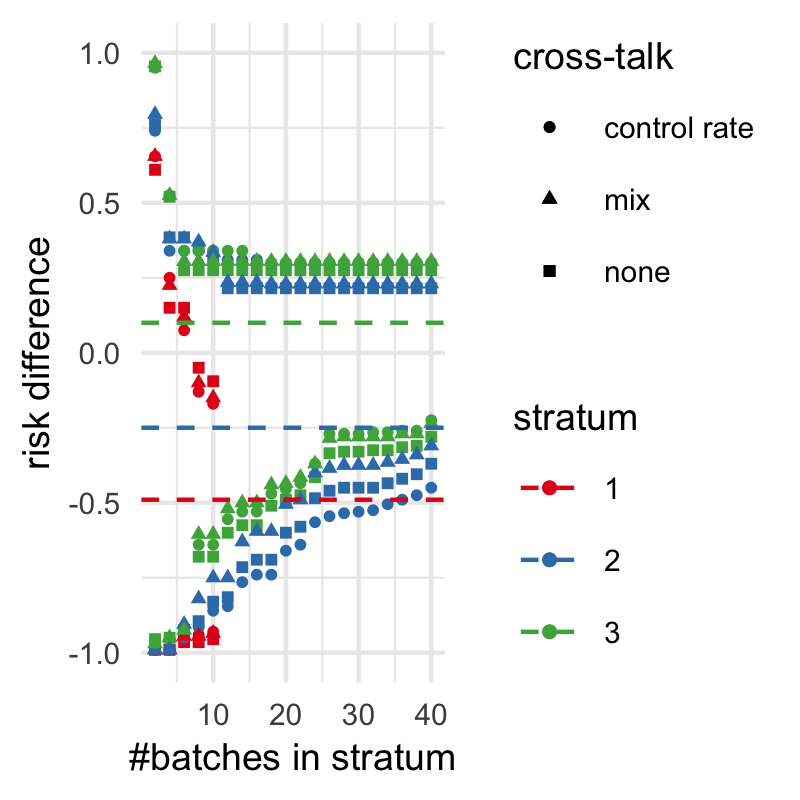}
         \caption{same control group rate}
     \end{subfigure}
     \hfill
     \begin{subfigure}[b]{4.5cm}
         \centering
         \includegraphics[width=\textwidth]{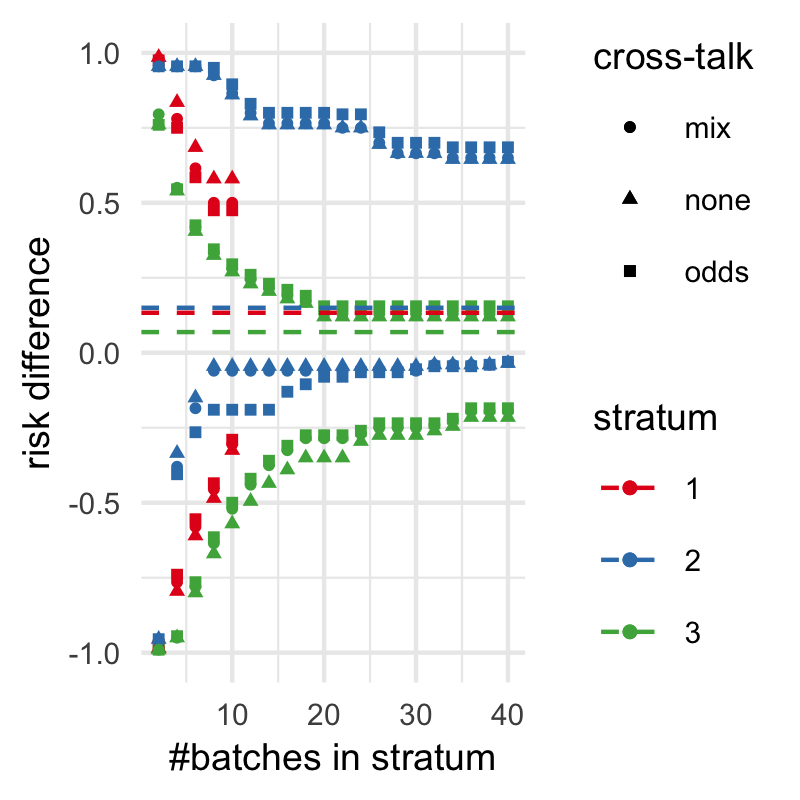}
         \caption{same OR}
     \end{subfigure}
     \begin{subfigure}[b]{4.5cm}
         \centering
         \includegraphics[width=\textwidth]{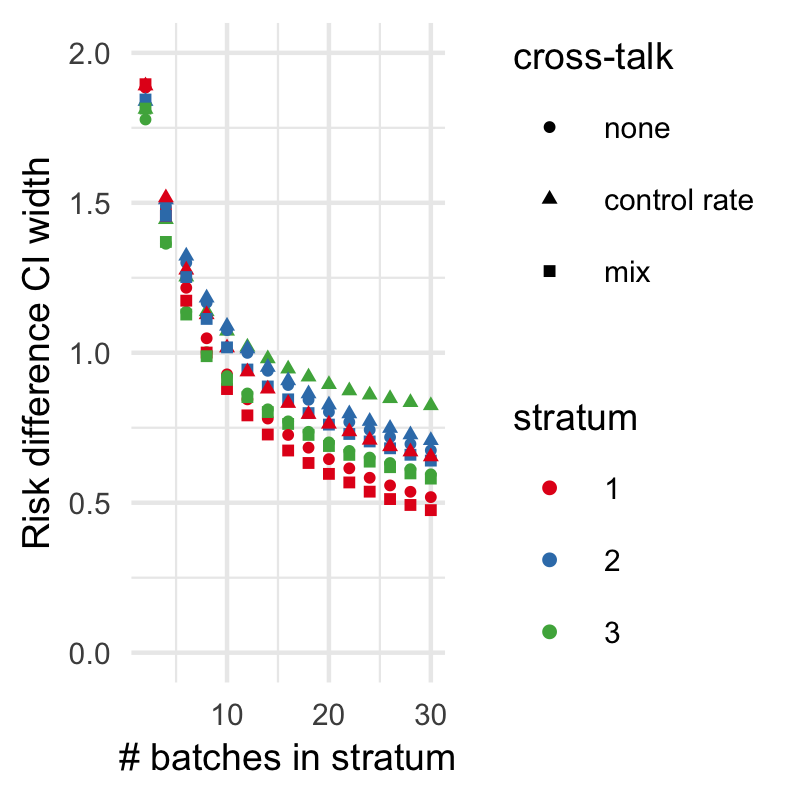}
         \caption{all different}
     \end{subfigure}
     \hfill
     \begin{subfigure}[b]{4.5cm}
         \centering
         \includegraphics[width=\textwidth]{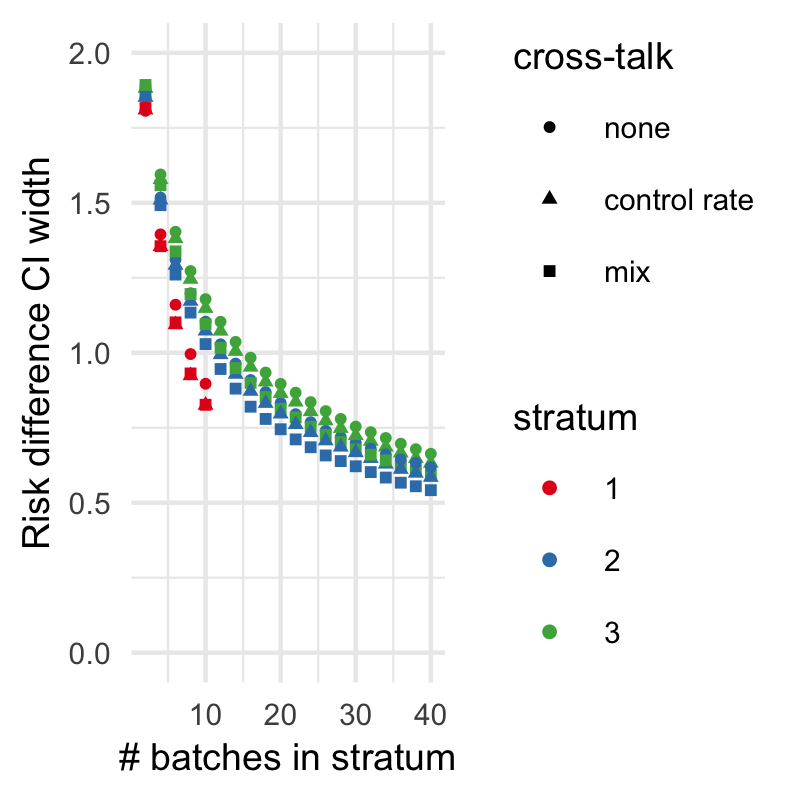}
         \caption{same control group rate}
     \end{subfigure}
     \hfill
     \begin{subfigure}[b]{4.5cm}
         \centering
         \includegraphics[width=\textwidth]{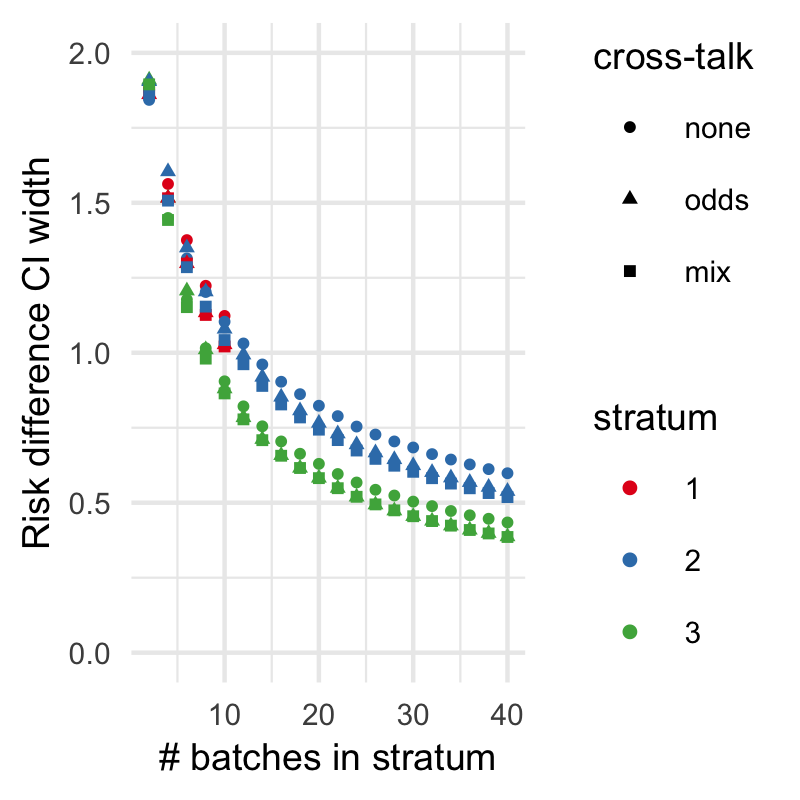}
         \caption{same OR}
     \end{subfigure}
     \vspace{.3in}
        \caption{Examples of $95$\% stratified confidence intervals ((a), (b) and (c)) and mean confidence interval widths estimated over $100$ runs ((d), (e) and (f)) with different types of cross-talk, including mixing different types of cross-talk. In (a), (b) and (c) the true risk difference of the data generating distribution in each stratum is indicated by a dashed line. For (a) and (d), the data were generated by distributions with different control group success rates ($0.1$, $0.2$ and $0.8$) and risk differences ($0.05$, $0.4$ and $-0.6$) in each stratum. For (b) and (e), strata sizes were unbalanced: as can be seen for stratum 1, the red points, data collection stopped after $10$ batches. Control group success rates were all $0.5$ and risk differences were different ($-0.49$, $-0.25$ and $0.1$). For (c) and (f), strata sizes were unbalanced as well, and now odds ratios were the same in each stratum ($2$), but control group rates differed again ($0.2$, $0.25$ and $0.85$).}
        \label{fig:crossTalkExamplesMix}
\end{figure}

\begin{figure}[htbp]
\vspace{.3in}
    \centering
    \begin{subfigure}[b]{4cm}
    \centering
    \includegraphics[width = \textwidth]{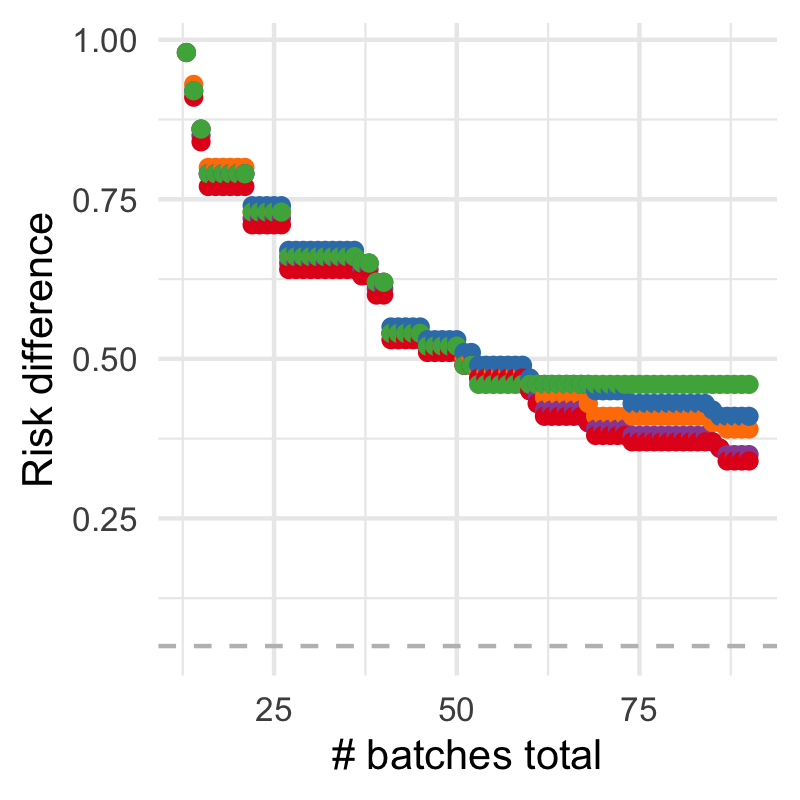}
    \caption{Upper bound sequence example}
    \end{subfigure}
    \hspace{1cm}
    \begin{subfigure}[b]{7cm}
    \centering
    \includegraphics[width = \textwidth]{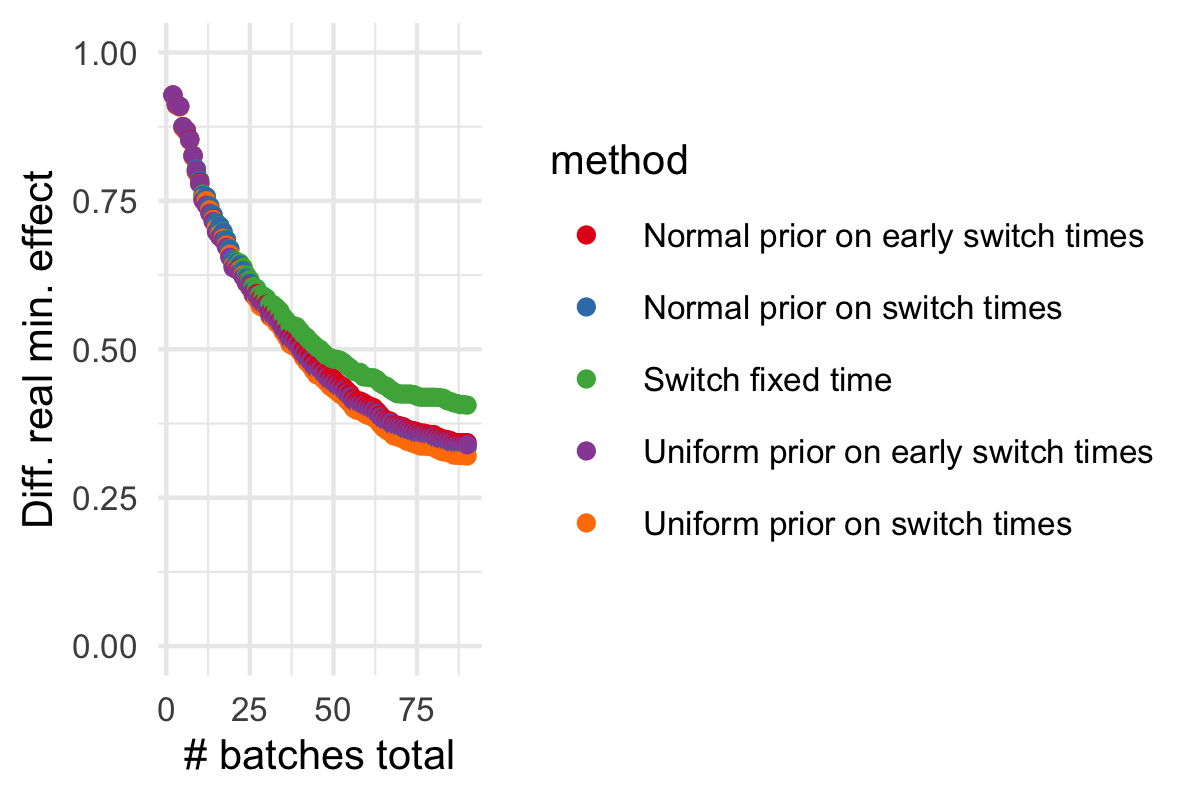}
    \caption{Average difference with true minimum}
    \end{subfigure}
    \vspace{.3in}
    \caption{Example of a confidence sequence and average difference from upper bound to true minimal effect size value through 100 simulations, for different switch priors on $j^*$. $30$ observations were made in each stratum, and the real differences were $0.5$, $0.4$ and $0.05$. For the priors on early switch times, all prior mass was distributed between batch numbers $5$ up to $10$.$\alpha$ was set to $0.05$.}
    \label{fig:switchPriorsBoundsMinDifference}
\end{figure}

\begin{figure}[htbp]
\vspace{.3in}
\centering
\includegraphics[width = 6cm]{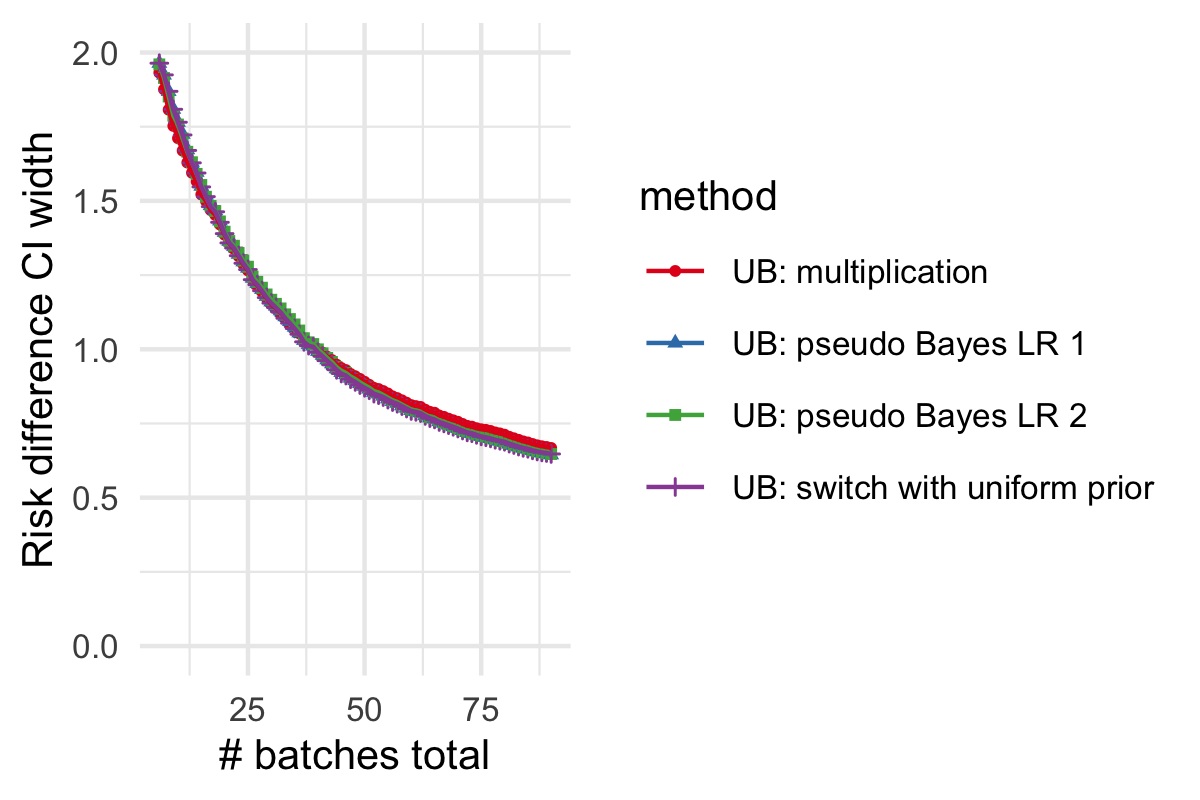}
\vspace{.3in}
\caption{Average interval width (upper bound for the respective methods minus lower bound estimated with the minimum method) of confidence sequences for the lower- (LB) and upper (UB) bounds of the minimum effect and estimated through 100 simulations. $30$ observations were made in each stratum, and the real differences were $0.5$, $0.4$ and $0.05$. With the switch method, a uniform prior ranging from $j^* = 5$ until $30$ was applied. With the pseudo-Bayesian approach, the learning rate $\eta$ was set to $1$ and $2$. $\alpha$ was set to $0.05$.}
\label{fig:CImindifferencsavwidth}
\end{figure}

\begin{figure}[htbp]
\vspace{.3in}
    \centering
    \begin{subfigure}[b]{6cm}
    \centering
    \includegraphics[width = \textwidth]{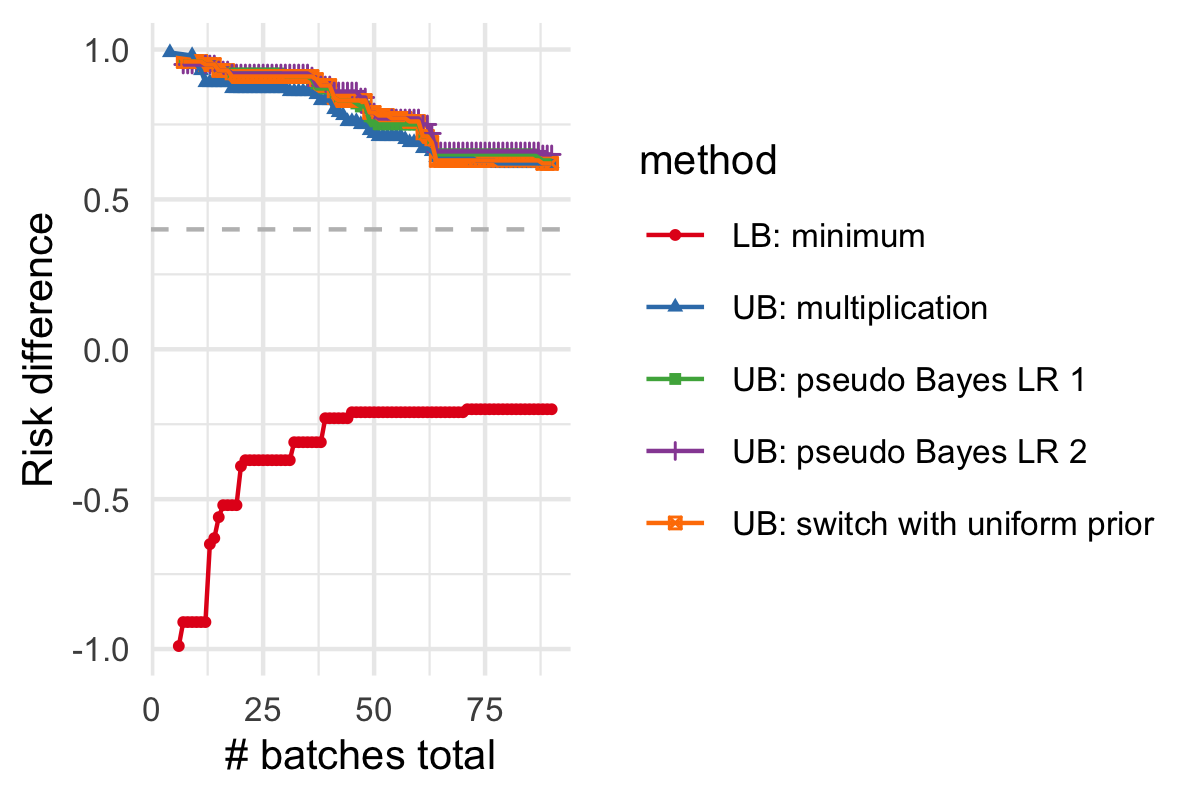}
    \caption{Confidence sequence example}
    \end{subfigure}
    \hspace{1cm}
    \begin{subfigure}[b]{6cm}
    \centering
    \includegraphics[width = \textwidth]{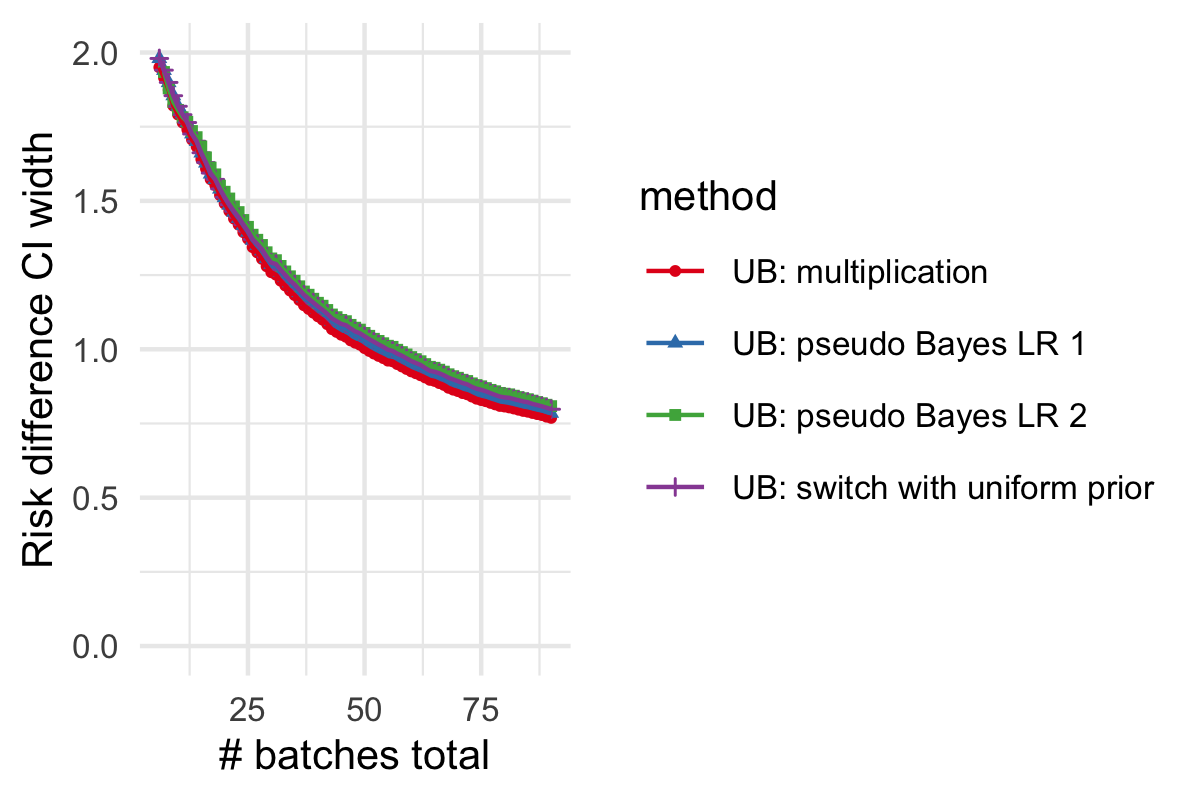}
    \caption{Average width}
    \end{subfigure}
    \vspace{.3in}
    \caption{Example of confidence sequences for the lower- (LB) and upper (UB) bounds of the minimum effect, and average interval width (upper bound for the respective methods minus lower bound estimated with the minimum method). $30$ observations were made in each stratum, and the real differences were $0.4$, $0.4$ and $0.5$. With the switch method, a uniform prior ranging from $m_{\text{switch}} = 5$ until $30$ was applied. With the pseudo-Bayesian approach, the learning rate $\eta$ was set to $1$ and $2$. $\alpha$ was set to $0.05$.}
    \label{fig:confIntervalForMinDifferenceEqual}
\end{figure}

\begin{figure}[htbp]
\vspace{.3in}
\centering
\includegraphics[width = 5cm]{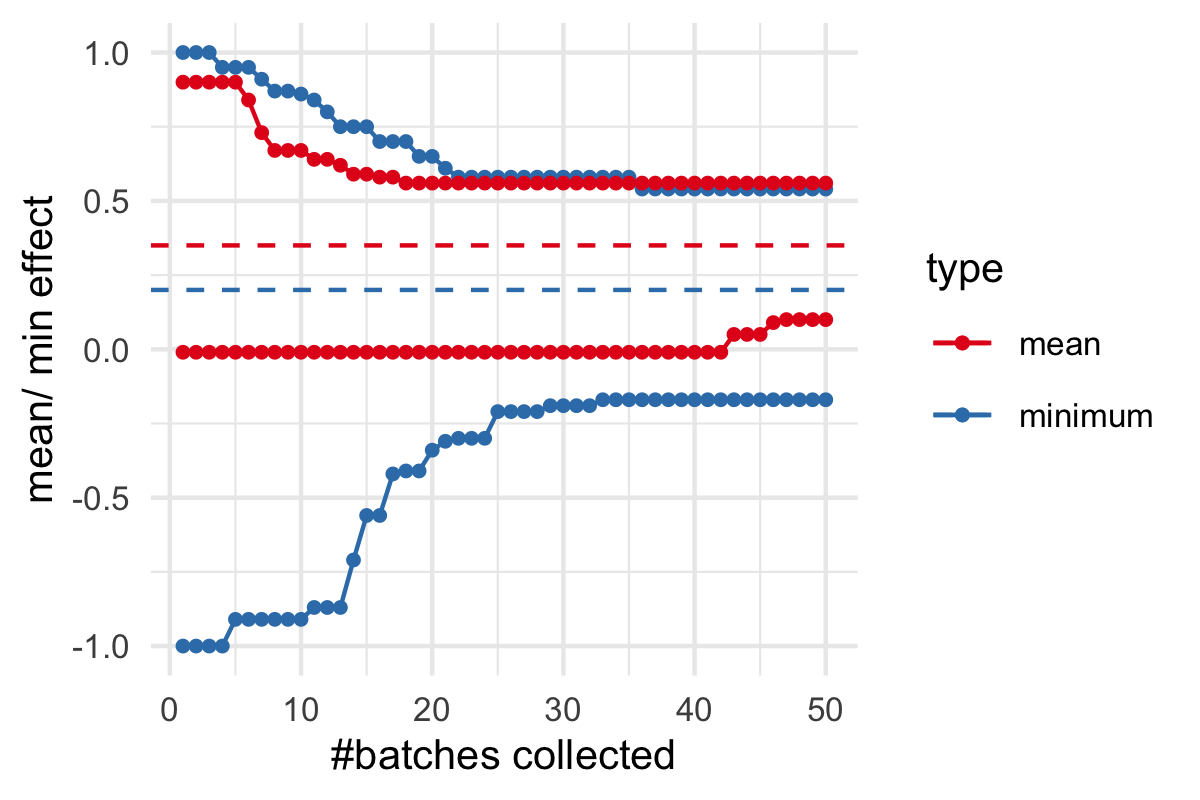}
\vspace{.3in}
\caption{Simulated example of a confidence sequence for
the mean effect across subpopulations. $25$ observations were made in each stratum, and the real risk differences were $0.2$ and $0.5$. The confidence sequence for the mean difference is plotted alongside the confidence sequence for the minimum of the differences, estimated with
pseudo-Bayesian averaging and a uniform switch prior. $\alpha$ was set to $0.05$.}
\label{meanvsminCI}
\end{figure}


\end{document}